\documentclass[12pt,preprint]{aastex}

\newcommand{\GN}{Gum Nebula}
\newcommand{\cgall}{CG~30/31/38\ }
\newcommand{\etal}{et al.\ }

\newcommand{\eg}{{\it e.g.}}

\newcommand{\sm}{M$_\odot$}
\newcommand{\Ha}{H$\alpha$\ }
\newcommand{\zpup}{$\zeta$~Pup}
\newcommand{\gvel}{$\gamma^2$~Vel}
\newcommand{\fxfbol}{$f_X/f_{bol}$}
\newcommand{\fxfv}{$f_X/f_{V}$\ }

\newcommand{\kmsec}{{km~s$^{-1}$}}
\newcommand{\EWLi}{{W$_\lambda$(Li~I)\ }}

\slugcomment{AJ in press}

\shorttitle{Star Formation in CG30/31/38}
\shortauthors{Kim et al.}

\begin{document}


\title{Low Mass Star Formation in the Gum Nebula: \\
      The CG~30/31/38 complex}


\author{Jinyoung Serena Kim\altaffilmark{1,2,3}, 
        Frederick M. Walter\altaffilmark{1} }
\email{serena@as.arizona.edu} 
\and
\author{Scott J. Wolk\altaffilmark{4}} 

%
\altaffiltext{1}{Dept. of Physics and Astronomy, SUNY at Stony Brook,
    Stony Brook, NY 11794-3800}
\altaffiltext{2}{Steward Observatory, 933 N. Cherry Ave., Tucson, AZ 85721-0065}
\altaffiltext{3}{Visiting Astronomer, Cerro Tololo Inter-American Observatory,
                 which is operated by the Association of Universities for Research
                 in Astronomy, Inc., under contact with the National Science Foundation}
\altaffiltext{4}{Harvard Smithsonian Center for Astrophysics, 60 Garden Street, 
Cambridge, MA 02138}


\begin{abstract}
We present photometric and spectroscopic results for the low mass 
pre-main sequence (PMS) stars with spectral types K -- M in 
the cometary globule (CG) 30/31/38 complex.  We obtained multi-object 
high resolution spectra for the targets selected as possible PMS stars 
from multi-wavelength photometry.  We identified 11 PMS stars brighter 
than V = 16.5 with ages $\lesssim$ 5 Myr at a distance of approximately 
200 pc.  The spatial distribution of the PMS stars, CG clouds, and 
ionizing sources (O stars and supernova remnants) suggests a possible 
triggered origin of the star formation in this region.  
We confirm the youth of the photometrically selected PMS stars using 
the lithium abundances.  The radial velocities of the low mass 
PMS stars are consistent with those of the cometary globules.
Most of the PMS stars show weak H$\alpha$ emission 
with W$_{\lambda}(H{\alpha})$ $<$ 10\AA.  Only 1 out of the 11 PMS stars 
shows a moderate near-IR excess, which suggests a short survival time 
(t $<$ 5~Myr) of circumstellar disks in this star forming environment.  
In addition, we find five young late type stars and one Ae star which
have no obvious relation to the CG~30/31/38 complex.
We also discuss a possible scenario of star formation history in the 
CG~30/31/38 region.  
\end{abstract}


\keywords{stars: formation  --- stars: pre-main sequence --- stars: low-mass ---
         (stars:) circumstellar matter-- (ISM:) HII regions --- ISM: globules}

\section{Introduction}

Most massive stars are found in their natal clusters and associations.
Low mass pre-main sequence (PMS) stars are also found in close proximity to 
massive stars, such as seen in Orion Nebula Cluster \citep{odel93,bal98}, 
and in vicinity of the Wolf-Rayet + O star binary system, \gvel\ \citep{poz00}.  
As soon as the massive stars begin to burn hydrogen, the protostellar clouds and 
T Tauri stars in the vicinity of the massive stars could be affected by the 
winds and strong radiation of massive stars.  In contrast to stars formed 
in such a strong radiation field environment, low mass stars formed in a quiet 
environment unaffected by massive stars, such as in the Taurus star 
forming region, may be able to retain larger reservoirs of circumstellar materials, 
which could prolong the lifetime of accretion disks.    
  
Globules are isolated dense clouds of molecular gas \citep{bok47} with masses of 
order of 10 -- 100 \sm\ and typical sizes of 0.1 -- 1~pc.  \citet{bok47} 
and \citet{bok77} initiated studies of star formation in globules, and these have 
been followed by many others \citep[and references therein]{sch77, rei83,
sug86, sug95, lef94, meg96, hes96, alv97, wol98, bal01, kim02, kim03a}.
These globules may form one to a few low mass stars in a very short time scale 
compared to the mean life time of globules, 1 -- few$\times$10$^6$ years, 
(see Elmergreen 1998 for review; and references therein).  
These globules are possible birth places for isolated low mass stars, 
such as the Sun.  
   
In this paper, we present a study of the star forming region \cgall, which lies
about 50 -- 100~pc from an early O star and OB associations.
This distance range is still close enough to be influenced by the radiation, 
but too far to be affected by the winds of the hot stars.  
The \cgall star forming region is a complex of small molecular cloud cores in 
the Gum Nebula evaporating due to ionizing radiation from an O star, $\zeta$~Pup,
and perhaps also due to progenitor of the Vela supernova remnant (SNR).  
The total ionizing radiation of two OB associations, Vela OB2 and Trumpler 10, 
may also be responsible for some ionizing fluxes in the Gum Nebula, and 
might have affected in this \cgall region.

\subsection{The Gum Nebula} 
The Gum Nebula, one of the largest H~II regions in our Galaxy, was discovered 
about half a century ago \citep{gum52, gum55}. It covers a large part of 
southern sky, in Vela, Puppis, Pyxis, Canis Major, and Carina.  It is located 
in the Galactic plane, with a 36$^\circ$ diameter, centered at 
$l=258^{\circ}$ and $b=-5^{\circ}$.  The nebula is the largest apparent 
H~II region, with a 36$^\circ$ diameter. The linear diameter is 250~pc at a 
distance of 450~pc \citep{bran71,rey76a,rey76b}.  The nature of the Gum Nebula 
remains unclear: this may be an old supernova remnant (SNR), a fossil 
Str\"omgren sphere \citep{bran71}, expanding H~II regions \citep{rey76a, 
rey76b, cha83} and/or a stellar wind bubble \citep[and references therein]{sah93}. 
The hot stars $\zeta$~Pup~(O4If) and $\gamma^2$~Vel~(WC8+O7.5I), and 
two OB associations (Tr~10 and Vela~OB2) near its geometric center 
can account for most of the photoionization in the nebula, and their
UV radiation probably is the cause of the evaporation of the 
CGs \citep{ber90, lef94}.  

The distances toward these associations, \zpup, \gvel, and the Vela SNR, and
their relation with the Gum Nebula are poorly understood, and remain controversial.  
The canonical values for the distances to these objects are about
400 -- 500~pc \citep[and references therein]{bran71, haw76, zea83, hen98}.
However, more recent studies, especially using the {\it Hipparcos} data suggest 
distances of 200 -- 400~pc \citep[and references therein]{dez99, knu00, hoo01}.

The Gum Nebula contains at least 32 cometary globules (CGs) 
\citep{haw76, san76, rei83, zea83}. Cometary globules are small bright rimmed 
isolated clouds with head and tail cometary morphology. These are evaporating 
clouds which are normally associated with H~II regions and OB stars.  
High resolution imaging with the {\it Hubble Space Telescope} has revealed 
smaller, solar-system size globules in Orion \citep{odel93, bal98} 
in which dusty protostellar disks are seen in silhouette.  Herbig-Haro (HH) objects 
and outflow sources are often found in Bok globules, cometary globules (CGs), and 
bright rimmed clouds.  Examples include B~335 \citep{fre82,kee83}, HH~120 in CG~30, 
HH~46/47 in the Gum Dark Cloud (GDC) 1 -- 7 \citep{bok77,rei83}, 
and CB~34 \citep{kee83,cle88,kha02}.  Some PMS stars have been identified
in vicinity of Bok globules and CGs \citep{rei83, pet94, alv97, kim02}.  

The tails of the CGs in the Gum Nebula are directed away from the central 
ionizing sources, $\zeta$~Pup, $\gamma^2$~Vel, and Vela pulsar, 
suggesting that CGs are evaporating away from a common center.  The locations 
of CGs and the direction of the tails show that CGs probably lie at the 
periphery of the wind-blown shell. If CGs were located on the line of sight 
between us and their ionizing sources, their tails would not be noticeable.  
However, the CGs located along the periphery of a sphere can be seen easily, 
because the tails evaporate in the transverse direction as projected in 
the sky. Therefore, even though CGs are located at about 50 -- 70 pc away 
from the ionizing sources, the dispersion in distances among the CGs can 
not be too large, for we are most likely to sample those near or along 
the periphery that show prominent tails.  

From his kinematic studies of the CGs in the Gum Nebula, Sridharan~(1992) 
showed that CGs are expanding away from the common center with a velocity 
of $\sim$12~\kmsec.  The velocity gradients along the CG tails imply that 
these tails have expanded for about 3 -- 6~Myrs.  An extended shell 
around Vela OB2 association, the IRAS-Vela Shell, has a size and location 
that seems to be consistent with the ring-like spatial distribution of 
the CGs \citep{sah92}.  Note that the radius of this ring of CGs are 
smaller than that of the Gum Nebula.  The expansion of this shell might 
have been caused by radiation driven stellar winds and multiple supernova 
explosions.  The expansion velocities of CGs and this shell, on 
the order of 10~\kmsec, are consistent with this, implying that the cause 
of CG expansion may be the same as the one for the IRAS-Vela shell. 

\subsection{The CG~30/31/38 complex}
\begin{figure}[t]
\centering
\includegraphics[totalheight=0.65\textheight]{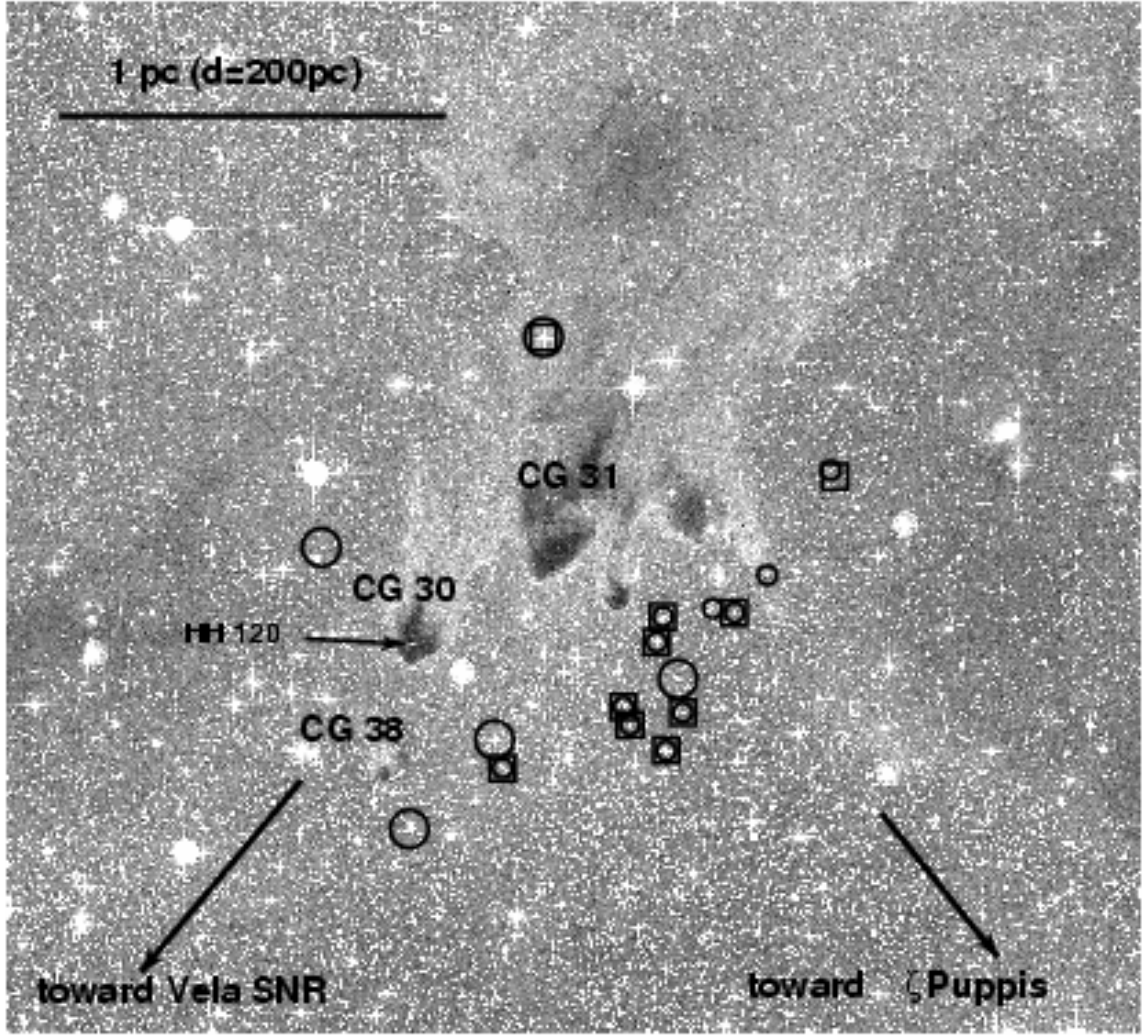}
\caption{The DSS image of the CG30/31/38 complex, showing the HRI sources
and the PMS stars. North is up; east is to the left.  
CG~30 has an outflow source HH~120; CG~31 is divided into 5 pieces, 
and CG38 is the smallest globule, to the southeast of CG~30.
The ten ROSAT HRI sources (boxes) appear to surround the heads of the CGs
on the side toward the central ionizing sources, while the CG tails point 
away from the ionizing sources.  
The small circles are centered on the spectroscopically confirmed PMS stars
that are associated with CGs, while larger circles are stars
that are less likely to be associated with CGs.}
\label{dss2}
\end{figure}

Cometary globules, such as the CG~30/31/38 complex in the \GN, 
provide a unique and convenient setting wherein to study star formation in 
an intermediate radiation field at a moderate distance from hot stars. 
Figure~\ref{dss2} is the Digital Sky Survey (DSS) image of the CG~30/31/38  
complex.  CG~38 is the smallest CG in this complex, seen just below CG~30.  
The head of CG~30 ($\sim$2\arcmin\ in diameter) contains a Herbig-Haro system,
HH~120, and the IR source CG~30-IRS~4 \citep{rei83,pet84}.  
Its $\sim$10\arcmin\ long tail points away from the central ionizing sources 
(i.e. Vela SNR, Vela~OB2).  CG~31 has 5 distinct clouds.
The tail of CG31~A is $\sim$25\arcmin\ -- 30\arcmin\ long (1 -- 1.5 pc at 200~pc).  
The change in the orientation of the tails of CG~31 coincides with the 
proper motion direction of $\zeta$~Pup.  Pettersson (1987) identified three 
early M stars with W$_{\lambda}$(H$_{\alpha}$)$>$10 \AA\
in this region as \Ha emitting T Tauri stars.

Recent studies of the reddening in this direction \citep{nie00,knu00} 
suggest a distance as close as 200~pc, about half the canonical value.
We use 200~pc as a conservative value for the distance of this cometary 
globule complex throughout this paper.  
\section{Observations}  
To identify PMS stars in the CG~30/31/38 region, we obtained  
X-ray images, optical, and near IR photometry.
We then obtained high resolution multi-object spectra of 
photometrically-selected samples to confirm the youth and membership of 
the samples. In this section, we describe these observations, the data 
reduction processes, and the calibration of the multi-wavelength photometry 
and high resolution multi-object spectroscopy.
\subsection{ROSAT/HRI photometry}
X-rays are an efficient means of selecting candidate low mass PMS stars
\citep[and references therein]{mon83, wal88}, because these stars are rapidly 
rotating and highly convective, the two ingredients which generate strong 
solar-like-coronal and chromospheric activity.
Magnetically active PMS stars (G, K, and M dwarfs) have X-ray luminosities 
of $>$10$^{29}$ erg/sec, and X-ray to bolometric flux ratio (\fxfbol) of about 
10$^{-3}$ -- 10$^{-4}$ \citep[and references therein]{wal88, fei02, fla02}.  
In contrast the Sun's X-ray luminosity varies from 10$^{26}$ to 10$^{27}$ erg/sec,
and its flux ratio varies from 10$^{-6}$ -- 10$^{-7}$.  
We use the spatial coincidence of bright stars with an X-ray-to-bolometric flux
ratio (\fxfbol) $>$ 10$^{-4}$ as the initial selection of the potential 
PMS stars.  Since this criterion also selects magnetically active foreground 
stars and fast rotators, we also obtained optical and near IR imaging 
data to select an unbiased photometric sample of candidate PMS stars. 

The CG~30/31/38 region was targeted in the ROSAT observation RH 201400, 
a 18364 second exposure obtained between 12 and 15 May 1994. 
The image was centered at $\alpha$=8$^h$9.2$^m$, $\delta$=$-$36$^{\circ}$6$^{'}$. 
Thirteen X-ray sources were detected in the 23\arcmin\ radius field of 
CG~30/31/38 region by the standard SASS source detection algorithm in 
12\arcsec$\times$12\arcsec\ or 24\arcsec$\times$24\arcsec\ detection cells 
(Table~\ref{Table_1}). Eight of the nine sources detected in the 
12\arcsec$\times$12\arcsec\ were recovered in the larger detection cell.  
These X-ray sources show a non-uniform spatial distribution, generally outlining 
the heads of the CG~30 and CG~31 (Figure~\ref{dss2}).

The ROSAT high-resolution imager (HRI) has little intrinsic energy resolution; 
we converted the HRI count rates (0.4 -- 2~keV) to fluxes using the standard 
energy-to-counts conversion factor (ECF)\footnote{see Table~8c and Figure~30 
from The ROSAT High Resolution Imager (HRI) Calibration Report,
ftp://sao-ftp.harvard.edu/pub/rosat/docs/hri\_report99/hri\_99.ps}
for Raymond-Smith spectral model. We assume a 1~kT thermal spectrum, and 
converted A$_V$ to N$_{\rm H}$ (N$_{\rm H}$=2.0$\times$10$^{21}$ cm$^{-2}$ mag$^{-1}~\times$ A$_V$).
We estimated A$_V$ values from the spectral types (see section 3.3) and 
observed $B-V$ colors, adopting intrinsic colors from dwarf stars 
\citep{ken95,luh99,leg01}. We assumed the standard galactic
extinction law with R=3.1.
%
\subsection{Optical followup photometry}
We obtained optical photometry of selected CG complexes in the Gum Nebula
on the nights of January 29 -- 30, 1996.
We used the CTIO 0.9m telescope with the 2048$\times$2048 Tektronix CCD 
(a pixel scale of 0\farcs4/pixel in a 13\farcm6 field of view)
and $UBVRI$ filters. 
During March 6 -- 12, 2001 we obtained more BVRI photometry in these fields 
with the same setup.  All data were 
obtained on photometric nights, with the mean seeing of 1\farcs2.

Bias, dome flat fields, and sky flat fields were taken at the beginning and
the end of each night.  Long (60 and 300 sec) and short (10 and 20 sec) 
exposures were taken for all the object fields to avoid saturation of bright 
stars in the fields. \citet{lan92} standard fields were observed several 
times per night to establish photometric calibration.  

Since the Gum Nebula lies in the Galactic plane there are typically more 
than 2000 stars in each field, many of which are heavily blended. 
We reduced the images with {\it IRAF/DAOPHOT} package.  
Since the point spread function (PSF) depends on the location within the CCD, 
we selected between 40 and 80 PSF stars in each field, and 
determined the spatial dependence of the
PSF using quadratic fits in both directions of the CCD.  

The full width half maximum values used for the {\it IRAF DAOPHOT} photometry
were between 1\farcs2 to 1\farcs8 pixels, while the radius for 
standard star photometry was 18 pixels.  Therefore, an aperture correction 
was made to the 18 pixel (7\farcs2) aperture.  The photometric errors are 
about 1\% at $V \leq 18$, and 5 -- 10\% at $V \sim 21$. 
The aperture correction is accurate to 1 -- 2\% in general.  

We used {\it imwcs} in wcstools \citep{min99} version 2.7.2 for the astrometric 
solutions. The positions of stars were determined using the Gaussian centroid 
algorithm in IRAF {\it phot} routine, then the astrometric solution was derived 
by fitting the pixel coordinates of stars to the known positions of USNO-A2.0 
astrometric standards (Monet 1998) located in each image. 
The {\it imwcs} package projects the image onto the tangent plane of the sky, 
rotates the image, and fits a polynomial to the x and y coordinates.  
In each image, about 120 -- 150 stars were matched with USNO-A2.0 astrometric 
standards with an average error of $\sim$0\farcs1.
We then averaged the positions measured in the V, R, and I band images for 
each star to give the final position. 

\subsection{Near-IR photometry}
We obtained JHK images covering all the X-ray sources using the CIRIM 
detector on the CTIO 1.5m telescope during February 5 -- 6, 1996 and 
February 17 -- 20, 1997.  CIRIM uses a 256$\times$256 HgCdTe NICMOS 3 array, 
and at f/13.5 gives a pixel scale of 0\farcs44/pixel.  Images were dithered by 
$\pm$15\arcsec\ in RA and DEC, and median filtered to determine level of 
the sky. For the targets we used a 6-point dither pattern.  
We used a 4--point dither pattern for the standard stars.

We reduced and analyzed the data using {\it the IDL/DoCIRIM} software
\citep{wal00} and custom written IDL codes.  We first generated median 
sky filtered and subtracted images.   Standard stars were chosen from 
the Elias (Elias 1982) and UKIRT FS standard star 
lists.\footnote{http://www.jach.hawaii.edu/UKIRT/astronomy/calib/phot\_cal/fs\_fundamental.html}
For astrometry we match stars in an image with USNO A2.0 catalog stars, 
and determine the image center and plate scale.  
We performed standard star photometry using a radius of 20 pixels (8\farcs8), 
and generated the photometric solution on the CIT system.
For object stars we used a default aperture radius of 3\farcs1, 
but we also used smaller apertures for objects that are 
faint or close to neighboring stars. We derived the aperture corrections from
the standard stars, and corrected our targets to 20 pixel apertures.
In this study, we used only those stars with photometric 
error $\le$5\%, which corresponds to K $\sim$15 -- 16 magnitudes.

\subsection{Matching optical, near-IR counterparts, and X-ray sources}
We matched the coordinates of X-ray sources with those of optical and 
near-IR stars to identify potential counterparts. The ROSAT/HRI X-ray 
source positions have data has much larger positional error
(see Table~\ref{Table_1}) than those of the optical and near-IR catalogs.
The matching algorithm, therefore, identified optical and near-IR sources
within the X-ray error circles (Table~\ref{Table_1}).
The HRI flux limit was such that a star matching the \fxfbol\ $>$10$^{-4}$
criterion has $V\lesssim$~16.
Since X-ray, optical, and near-IR data were not obtained simultaneously, 
and PMS stars are variable, we also considered stars up to 1~mag fainter.
Twelve of the HRI sources have bright ($V\lesssim$ 17; $K\lesssim$ 13)
optical and near-IR counterparts within the 3$\sigma$ X-ray error circles.
For the X-ray sources that were matched with more than one optical and IR source, 
we chose the brightest sources among all the sources within the error circle.
The faintest optical counterpart, XRS~3, has $V$ magnitude of 17.27; 
XRS~1 does not have an optical counterpart with $V\lesssim$~17 within its 
$3\sigma$ X-ray error circle.

\subsection{Optical spectroscopy}
During March 6 -- 8, 2002 we obtained multi-object high resolution spectra of 
the candidate PMS stars (see Section 3), using the HYDRA multi-fiber 
spectrograph on the CTIO Blanco 4m telescope in echelle mode.
The goals of these spectroscopic observations were to confirm the youth 
of these stars using Li~I 6708\AA\ absorption lines and to measure their
radial velocities. The spectra also reveal the spectral types, and we can 
estimate extinction toward the PMS stars by comparing the measured colors 
with the intrinsic colors expected for the observed spectral types.  
The radial velocities let us determine membership probabilities, by
comparing the distribution of velocities with that of the CGs. 
Using CTIO/HYDRA, one can obtain spectra of up to 135 objects 
(stars and sky) in a 40\arcmin\ field of view simultaneously.  Since the overall 
efficiency of the HYDRA in echelle mode is about 3\%, we chose targets that 
were brighter than $V \sim 17$ mag.  The S/N was sufficient 
to detect Li~I lines with strengths expected of PMS stars.

\subsubsection{Spectroscopic sample selection}
We selected as targets those stars that are coincident with the HRI X-ray 
sources and other stars whose colors lie in the appropriate region in 
the Color-Magnitude Diagram (CMD) for PMS stars at a distance of $\sim$200~pc.   
$V$ magnitudes of targets range between 10 and 17.
In each field there are between 50 and 200 objects near the PMS locus in the CMD 
(Figure~\ref{cmd_phot}).  At the bluer end ($V-I \sim$1.0) of the CMD, 
the non-X-ray emitting stars in the PMS locus 
are heavily contaminated by background stars in the galaxy, 
but the fainter and redder candidates are less likely to be contaminated. 
Since the Gum Nebula is located in the Galactic plane, some of the fields
suffer severe contamination.

\subsubsection{Spectroscopic observations, data reduction, and wavelength calibration}
We observed  144 objects in two 40\arcmin\ diameter fields on March 6 and 7, 2002. 
Three exposures were taken of each target field.
These were combined using the IRAF {\it combine} with the median sigma clipping
algorithm.  Three offset sky spectra were also taken at a position 
offset 10\arcsec\ from the targets after each target exposure.
These sky spectra have 10\% of the integration time of the target spectra.  
These sky offset images were used to identify background nebula emission 
line profiles.  The sky subtraction is discussed in more detail in section 2.5.3. 

We used the IRAF/HYDRA package for basic data reduction, and also some custom 
written IDL codes for various calibration purposes.  We first fit the CCD 
overscan of the each image interactively, and trimmed all images.
We then created a combined bias (zero) for each night, and subtracted 
it from all images. We created a {\it milky flat} image from daylight 
sky exposures, which we used to flatten all other images.

Scattered light was subtracted from images using {\it abscat} task.
The scattered light is a global 2-D feature, and is subtracted 
before each aperture is traced and extracted.  Then the {\it IRAF/apall} task 
was used to extract spectra.   The FWHM of a fiber is about 6 -- 7 pixels, 
so we selected 3 pixels for upper and lower widths, and extracted a total of 
6 pixels centered at the center of each trace.

We took about 5 dusk and 5 dawn twilight sky spectra each night to make combined 
dusk and dawn sky spectra.  
Th-Ar comparison lamp exposures were taken at the beginning and end
of each night for wavelength calibration.  The mean residual of the wavelength
solutions is $\leq 0.08$\AA\ (0.5 pixel).   
Etalon comparison lamp exposures were taken before and after taking Th-Ar lamps, 
and between every configuration of object fields to correct for the zero point 
shift throughout the night.  We selected as reference the 
etalon spectrum which was taken immediately before or 
after the Th-Ar comparison exposure each night; all other etalon spectra were 
cross-correlated in pixel space.  We calculated the zero point shifts in pixels, 
and then applied the shift to derive wavelength vectors for all object spectra.

Since the Th-Ar signal was weak during the first night of observation,
because of a faulty lamp, we used the second night's Th-Ar solution.  
Some apertures that contained high enough S/N in first night's 
Th-Ar spectra were cross-correlated with the second night's Th-Ar spectra. 
The median pixel shift of 0.023 pixels is very small.
We applied this shift to first night's Th-Ar solution;
the rest of the wavelength calibration using the etalon was done
the same way as on night~2.  
{\it IRAF/dispcor} was then used to dispersion-correct sky spectra and object
spectra using Th-Ar solution as reference spectrum.  The starting wavelength
and dispersion (\AA/pix) were derived for each of the 135 fibers; 
then the zero point was shifted using the cross-correlation results from 
Th-Ar and etalon spectra.
%
\subsubsection{Sky Background Subtraction}

The nebulosity in these regions complicates the sky subtraction.
In addition to the offset sky spectra, we also devoted
between 10 and 20 fibers in each configuration field to sky.  Here we use the 
term {\it sky background} to refer to the emission lines due to nebulosity, 
which is particularly important for measuring the strength of the stellar \Ha 
lines. The sky emission line strengths of \Ha and [N~II] vary across the 
40~\arcmin\ field of view.  The background sky line profiles vary spatially due 
to the kinematics of the expanding gas shell in the Gum Nebula. 

We group the sky spectra into three general categories, based on the \Ha\
and [N~II] line profiles and strength.  Group~1 has narrow and strong \Ha 
and [N~II]; group 2 has  asymmetric and broader line widths than group 1; 
and group 3 has broader line widths than group 2, possibly with 
double-peaked lines. 

To check the strength and profile of night sky emission lines, we compared 
the night sky spectra obtained during the long exposure with the combined 
three offset sky spectra which were taken 10\arcsec\ offset from the target 
field immediately following each of the three sub-exposures.  Although the 
strength of the \Ha lines in the offset sky spectra were 10\%  of long 
exposure sky lines, the offset sky spectra helped determining which subgroup 
of sky profile should be used for each star. 

We subtracted the appropriate sky spectrum from the object spectra. 
Due to the spatially variable background sky line profiles, sky background 
can not be perfectly subtracted.  However after careful checking, we are 
confident that we can derive the equivalent width of \Ha to an accuracy 
of $\sim$0.8\AA.  Line emission is not a problem at the $\lambda$6708~\AA\ 
wavelength of Li~I.  
%
\section{Results}
From X-ray, optical, near-IR photometry, and optical spectroscopy, 
we have identified 16 young stars (Table~\ref{tbl:EW}).  
Eight of the 13 X-ray sources (XRS) listed in the Table~\ref{Table_1} 
are spectroscopically confirmed young stars, 
two X-ray sources were not observed, and three do not appear to be young stars.
We identified eight more young stars, not detected in the ROSAT/HRI
observations, based on the Li~I $\lambda$6708 \AA\ absorption line.  
Of the 16 stars, radial velocities of eleven are consistent with 
that of the CGs \citep{sri92, nie98}.  In the next section we discuss our 
results from X-ray photometry; optical and near IR photometry; and optical 
spectroscopy.

\subsection{X-ray photometry}  
In Table~\ref{Table_1} we present X-ray count rates of the 13 X-ray sources. 
We also present the offsets in X-ray coordinates and the optical counterparts. 
Table~\ref{tbl:EW} lists spectroscopically confirmed young stars along
with their spectral types, \Ha and Li equivalent widths, Li abundances, 
and A$_V$ values. 

Using the ROSAT/HRI photometry, optical photometry and spectroscopy, 
we derived X-ray to bolometric flux ratio for 10 X-ray sources 
(Table~\ref{Table_1}).  
Using the spectral types (see section 3.3.1), we adopted bolometric 
correction values from a color table that was compiled using 
published color tables by \citet{ken95,luh99,leg01}, which can be
found in Table~3 of \citet{sher04}.  

We used Table~3 from \citet{sher04} to estimate unreddened colors and
bolometric corrections for our spectroscopic sample.  This table combines
spectral types, bolometric corrections, and colors from \citet{ken95} 
and \citet{leg01} with the PMS temperature scale of \citet{luh99}.  
We derived A$_V$ for each star by comparing its predicted unreddenned 
B -- V color to the observed B -- V color.  As a check on the derived 
A$_V$s, we compared the spectral energy distribution (SED) for each star 
derived from our V, R, I, J, and H band data to model SEDs for K and M 
stars with a range of A$_V$s.  A$_V$s derived from the SEDs were consistent 
with the A$_V$s we calculated from the B -- V colors, but with larger 
uncertainties on A$_V$.  This suggested that our uncertainties on 
A$_V$ were two -- three times as large as we had estimated from the 
B -- V colors. The larger uncertainties are incorporated into 
Table~\ref{tbl:EW} and Figures~\ref{fxfbol} and \ref{fxfbolTeff}.

\subsubsection{\fxfv\ and \fxfbol}
Figure~\ref{fxfbol} shows X-ray flux vs. bolometric flux ratio of 10
X-ray sources listed in Table~\ref{Table_1} in logarithmic scale 
($\log(f_X)$ vs. $\log(f_{bol})$).  XRS~5 and XRS~12 are confirmed 
not to be PMS stars, and since we do not have optical flux of XRS~9,
we do not include these three sources in this plot.
Two high $f_X$ points are KWW~1863 (XRS~4) and KWW~1637 (XRS~6). 
These high count rates suggest strong coronal activity or flaring.  
All the sources appear above $log$(\fxfbol) = -3 line.

Figure~\ref{fxfbolTeff} is a X-ray to bolometric flux ratio vs. 
effective temperature in logarithmic scale ($\log$(\fxfbol) vs. 
$\log$(T$_{eff}$) plot of the 10 X-ray sources plotted in Figure~\ref{fxfbol}.  
The $\log$(\fxfbol) values ranges from $\sim$-3 up to $\sim$-1.5. 
Active K and M type stars are known to have $\log$(\fxfbol) ratio 
from $-$4 up to $-$1, while  $\log$(\fxfbol) of PMS stars with
spectral types earlier than G type range between -4 to -2.

In Figure~\ref{fxfbolTeff} we overplot 3$\sigma$ detection limits.  
These stars lie just at or slightly above this 3$\sigma$ detection limit.  
The mean $\log$(\fxfbol) for other star forming regions is about $-$3.5 
\citep[and references therein]{fei02,fla02,fla03}.  This suggests that the 
HRI sources represent $\lesssim$15\% of the X-ray bright PMS population if 
we assume that $L_X/L_{bol}$ of the 0.1 $-$ 2 \sm\ PMS stars follows that of 
typical young star clusters.  If the log($f_X/f_{bol}$) of this star forming 
region follows the distribution of ONC \citep{fla02,fla03}, for the mass 
range of 0.2~\sm\ -- 1.0~\sm, there could be roughly six times 
more PMS stars with log($f_X/f_{bol}$)~$<$~$-$3.0. 

\subsubsection{Variability}
We examined the X-ray light curves for variability.  Since the number 
of source counts is small, we could look only for evidence of gross changes 
in the count rate.  Of the brighter sources, only KWW~1637 (XRS~6), a 
double-lined spectroscopic binary PMS system, shows any evidence of variability, 
at about the 2$\sigma$ confidence level, in 30~ksec bins. 
The variation on timescales between 30 and 250~ksec is no more than 
about 30\% from the mean.  Among the weaker sources, KWW~1055 (XRS~10) 
appears variable.  All the counts arrived in the first half of the observation. 
The probability that the source is constant is 0.007 based on $\chi^2$ statistics.
\subsection{Color-magnitude diagram} 
\begin{figure}[p]
\plotone{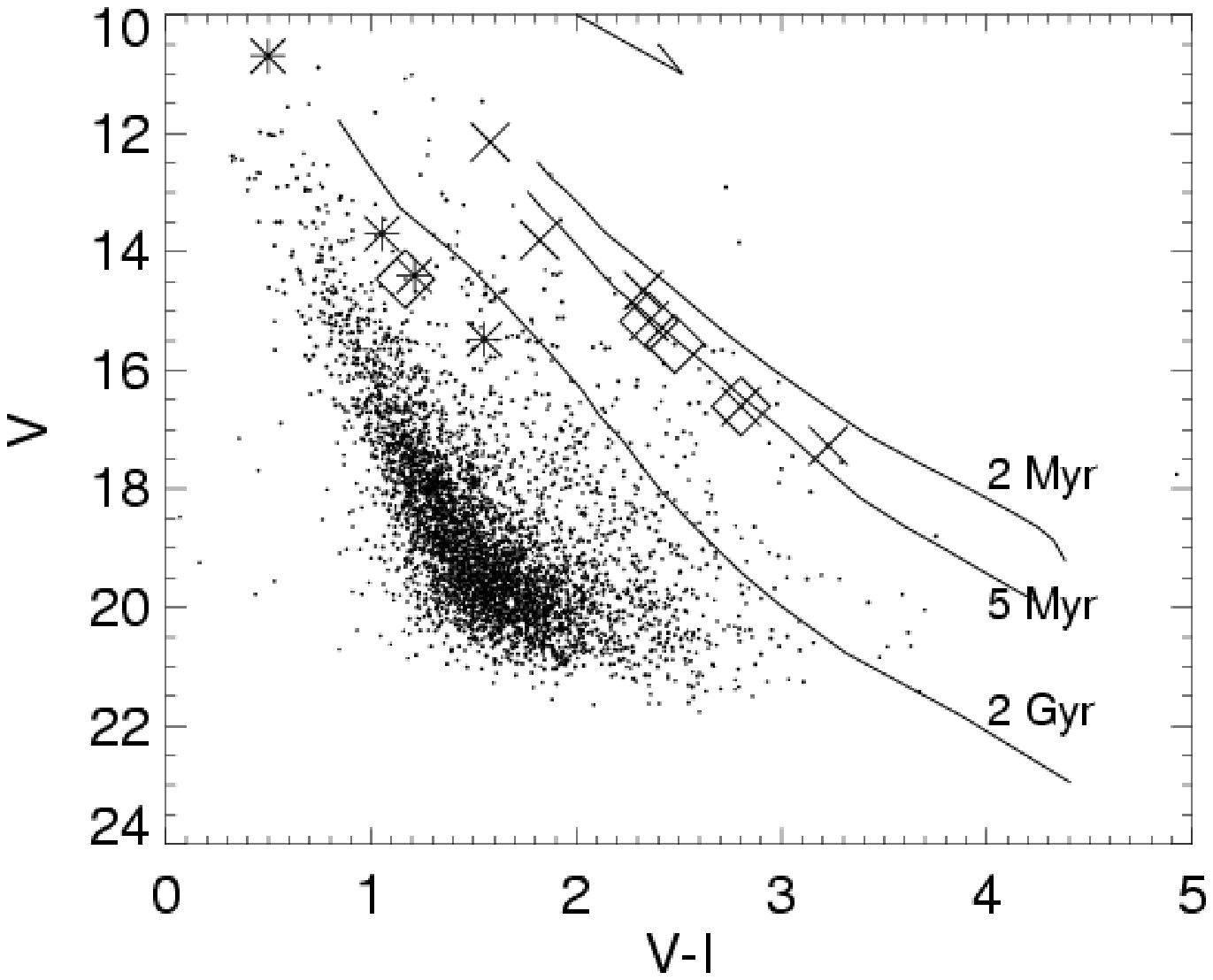}
\caption{
 The $V$, $V-I$ color magnitude diagram (CMD) of CG~30/31/38 complex.
 Stars which are likely counterparts of X-ray sources are indicated with
 crosses and asterisks.  Diamonds mark \Ha emission sources
 \citep{pet87,her88,sch90}.
 The A$_V$=1 mag reddening vector, plotted at top center, runs almost parallel 
 to the isochrones.  Isochrones are from Baraffe \etal (1998) using color
 tables by \citet{ken95,luh99,leg01} for 2~Myr, 5~Myr, and 2~Gyr
 at an assumed distance of 200~pc. We use the evolutionary models with
 [M/H]=0 and y=0.275.
}
\label{cmd_phot}
\end{figure}

\begin{figure}[t]
\plotone{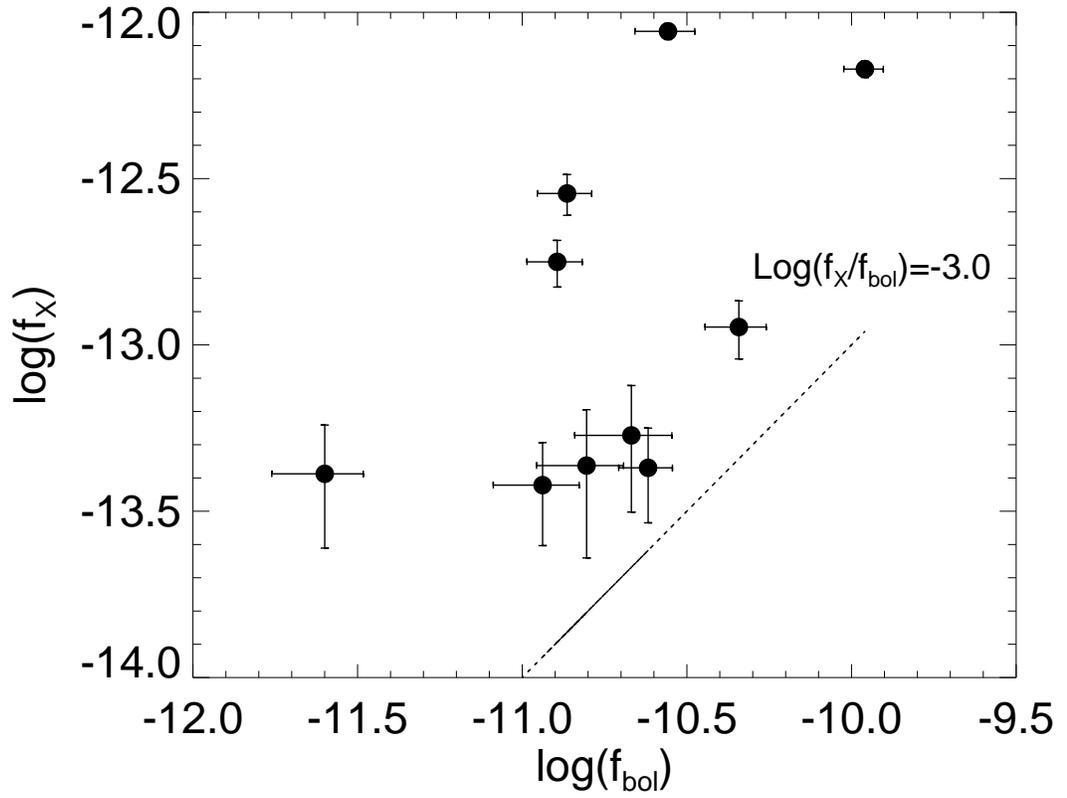}
\caption{
The $\log f_X$, $\log f_{bol}$ plot for ROSAT/HRI sources. 
The dotted line is drawn at $\log$(\fxfbol) = $-$3.}
\label{fxfbol}
\end{figure}

\begin{figure}[p]
\plotone{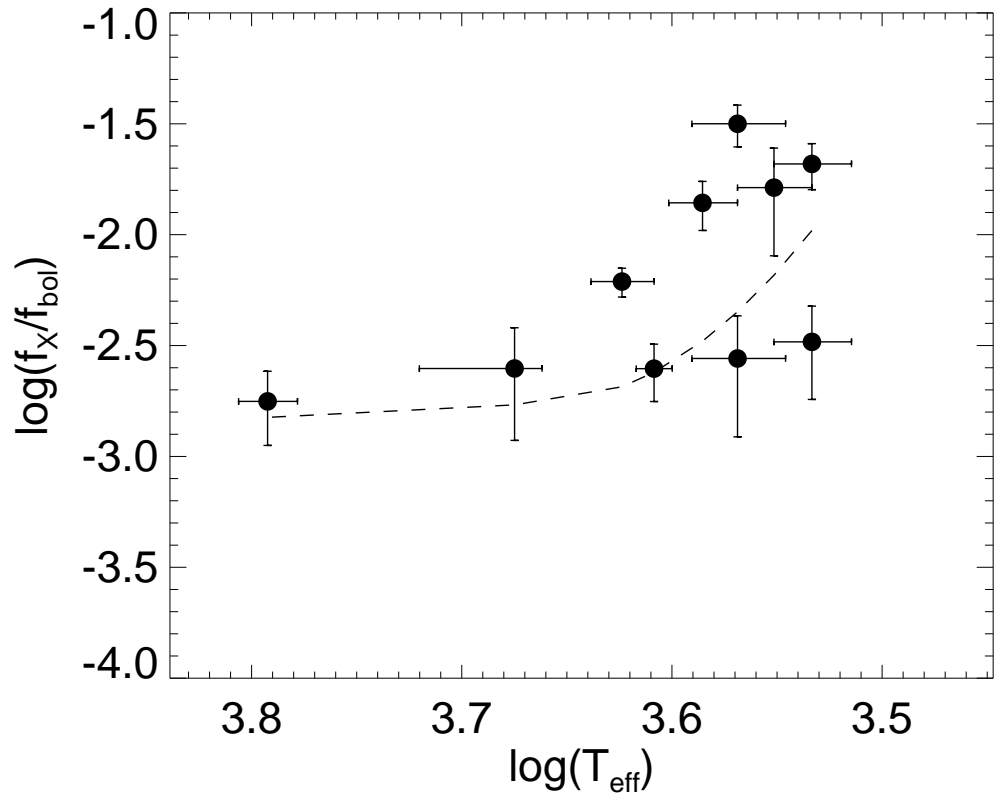}
\caption{
$\log$(\fxfbol) plotted as a function of T$_{eff}$. The dashed line shows 
the approximate 3$\sigma$ detection limit. Half of the X-ray sources are 
on or slightly over the sensitivity limit, which suggests that there are
more PMS stars awaiting to be discovered.
}
\label{fxfbolTeff}
\end{figure}

Figure~\ref{cmd_phot} is the $V, V-I$ CMD for objects in the CG~30/31/38 region. 
Stars that are likely counterparts of X-ray sources are indicated with crosses 
and asterisks.  The limiting magnitude of X-ray flux, assuming log(\fxfbol) = -4, 
corresponds to $V$ magnitude of $\sim$16. 
The reddening vector, plotted on upper part of the CMD, is a line running almost 
parallel to the isochrones. Evolutionary tracks and isochrones are from 
\citet{bar98} modified using temperature scale and colors of 
\citet{ken95,luh99,leg01} for low mass stars.  Isochrones for ages of 
2~Myr, 5~Myr, and 2~Gyr are plotted for a distance of 200~pc. 

The positions X-ray sources plotted in the CMD (Figure~\ref{cmd_phot}) seem 
to delineate two distinct loci parallel to either the ZAMS or a PMS isochrone.
The upper locus is defined by six X-ray sources ($\times$ symbols) and three 
previously cataloged PMS stars (diamond symbols; \citet{her88}). 
A  5~Myr isochrone \citep{bar98} at the 200~pc distance of the cloud complex 
is a good match to this locus (see Figure~\ref{cmd_phot}). 
Four X-ray sources (asterisks) and one \Ha source (diamond; \citet{sch90})
trace a distinctly different locus about 2 -- 3 magnitudes fainter than the
upper locus.  At a 200~pc distance these objects lie below the main sequence; 
they must represent a distinct and more distant population.
 
By using the minimum likely distance of 200~pc to CG~30/31/38 complex 
\citep{knu99}, we can place an upper limit on the age of the stars if they 
are indeed PMS stars at a common distance.
The uncertainty in the age does correlate with the uncertainty in the distance.
However, for young low mass stars one may use Lithium abundance to further 
constrain the age (see \S3.4).
For example, M stars with log~N(Li) $\sim$3 are likely to be younger than 
$\sim$5~Myrs old \citep[and references therein]{wal94,mar94}.
Conversely, by assuming (or measuring) an age we can constrain the distance
to the cloud complex, if spatially related to the stars.
\subsection{Spectra} 
Multi-wavelength photometry enables us to produce a large and unbiased 
sample of candidate PMS stars, but is subject to contamination.  
Because photometry alone can be ambiguous, spectroscopy is necessary to 
confirm the youth and membership of stars using Li~I $\lambda$6708~\AA\ lines 
and radial velocities.  In the previous sections, we described selection of 
candidate PMS stars using X-ray, optical, and near IR photometry.  
To confirm the youth of our photometrically selected PMS stars, 
we obtained high resolution multi-object spectra of our candidates. 
In section 2.5 we described the multi-fiber spectroscopic observations using 
CTIO-4m/HYDRA, data reduction, and calibration.  Here we present the spectra 
and results from our spectroscopic observations and analysis.

\subsubsection{Spectral type determination}  
Spectra of PMS stars are shown in Figures~\ref{allspectra-a} --
\ref{allspectra-e}.
We flattened the continua 
of the spectra using a polynomial fit, and normalized to unity. 
We estimated the spectral types by comparing with spectra of 26 standard 
stars obtained at comparable dispersions.

The $\lambda\lambda$ 6400 -- 6720~\AA\ wavelength range is not ideal for 
spectral type determination, because of the paucity of strong lines that 
can be used for spectral type determination.  We estimated the 
spectral types mainly using Fe~I and Ca~I lines. 
The strength and structure of the $\lambda$6495~\AA\ blend was used to
distinguish spectral types for spectral range of F -- K. 
TiO bands were used for the determination of late K and M spectral types.

The results are summarized in Table~\ref{tbl:EW}, and also discussed 
in the Appendix. The uncertainties on the spectral type are about 0.5 
subtypes for K and M stars, and are larger for stars with 
earlier spectral types.  

\subsubsection{Spectroscopically confirmed PMS stars} 
In Figures~\ref{allspectra-a} -- \ref{allspectra-e} we show the spectra of 
the confirmed and possible PMS stars.  Two of the confirmed 
PMS stars (KWW~1892 and KWW~1953) were observed twice.
Three of the X-ray sources (XRS 5, 11, and 12) in the photometrically
selected sample do not appear to be PMS stars.  Spectra of KWW~1043 
(XRS~8) and XRS~13 were not obtained, because these were too faint. 
We identified the Li~I 6708~\AA\ absorption line in 15 stars. Ten of 
these also have \Ha in emission.  Only KWW~873 shows near-IR excess.
One star on the lower locus, KWW~314, is an early type star with an \Ha 
profile reminiscent of an Ae star. The spectral type is A3 (see the
Appendix). Not surprisingly, no Li absorption is evident.

Based primarily on the presence of the Li~I 6708~\AA\ line (next section),
we confirm that eight X-ray sources. KWW numbers 464, 1892, 598, 1863, 
1637, 873, 1055, and XRS~9, are PMS stars. 
\Ha line profiles of these PMS stars are described in more detail in the
Appendix.  We also find 8 other young or active stars that were not detected 
(or not observed) in X-rays, KWW~314, 975, 1125, 1302, 1333, 1806, 1953, 
and 2205 (see Table~\ref{tbl:EW}). 

Three PMS stars from the Herbig-Bell catalog (Herbig \& Bell 1988),
HBC 553, 554, and 555, also lie along the upper locus in the 
$V$, $V-I$ CMD (diamonds in Figure~\ref{cmd_spec}).
We observed two of these stars, KWW~1892 (HBC~553) and KWW~975 (HBC~554).
We did not observe KWW~1043 (XRS~8), the likely counterpart of HBC~555.
Table~\ref{tbl:EW} includes the derived properties of these HBC sources.  

Short descriptions of each of these stars can be found in Appendix. 
Table~\ref{tbl:EW} summarizes our results.  
In Figure~\ref{cmd_spec} the stars marked with $\times$'s and asterisks are 
the X-ray sources as in Figure~\ref{cmd_phot}.  We mark the confirmed PMS stars
with filled circles (upper locus) and open triangles (lower locus).
Note that the PMS stars in the upper locus are clustered spatially below
the CG~31 complex (boxes with circles in Figure~\ref{dss2}).

Inclusion of the optically-identified young stars strengthens the case for
the existence of 2 parallel loci in this CMD (Figure~\ref{cmd_spec})
The upper locus is idefined by the X-ray sources ($\times$), HBC sources 
(diamonds), and spectroscopically-young stars (filled circles).  
The Li abundance, \fxfbol\ ratio, and the \Ha emission, are all consistent 
with the interpretation of the upper locus as an isochrone defining a PMS 
stellar population.

The lower locus is defined by X-ray sources (asterisks) and 
spectroscopically confirmed young stars (open triangles).
Note that only one X-ray source (KWW~1055) is confirmed to be a PMS star,
while XRS~11 and 12 are not PMS stars.  As mentioned earlier, the spectrum 
of XRS~13 was not obtained. If the stars on the lower locus are at 400~pc, 
the age inferred from isochrone fitting is $>$100~Myr. 
If young, these stars are likely to be at a distance greater than 400~pc. 
They may be unrelated to the CG complex and the stars on the upper locus.

\subsection{Lithium abundance}  
The lithium abundance is of interest because it both probes stellar 
structure and age, and also constrains primordial nucleosynthesis models 
\citep[and references therein]{dun91}.
The universal primordial Li abundance, logN(Li), is 3.1 
(on the scale of logN(H)=12), however the Li abundance of halo stars
is 2.1, while the mean distribution of logN(Li) values for PMS stars 
in young clusters peaks at 3.1 \citep{dun91,mar94}. 

High Li abundances can be indicative of the youth of solar mass 
or less massive stars, because the temperature at the base of convection 
zone is still too low to burn lithium for 1 -- 10 Myrs.  
These cool stars (M~$\leq$~1~M$_\odot$) have deeper convective envelopes, 
and processes such as convective overshooting mix the surface and deeper 
(hotter) layers causing faster Li depletion than in higher mass stars.
For higher mass stars (M~$>$~1~M$_\odot$) Li depletion is not significant, 
because they spend less time in the PMS stage, and the convection zone 
is thinner than in low mass stars \citep{king93,ven98}.
Because of this fast lithium depletion, lithium can be a good indicator 
of ages in low mass stars.  

We derive lithium abundances, log~N(Li) using both NLTE \citep{pav96} 
and LTE \citep{dun91} analyses.  
The results are presented in Table~\ref{tbl:EW}.  
Since these curves of growth are valid for T$_{eff}~>$~3500~K,
the curve of growth for lower temperature were extrapolated for
the stars 3200~K~$<$~T$_{eff}~<$~3500~K.
The derived Li abundances for K6 and K7 stars (XRS~6 and XRS~7) are higher
than the universal value (3.82 and 3.56 for NLTE). 
The high value for these K stars are consistent with observations
of other PMS stars, \eg, \citet{bas91, mar94, wal97b}.

The individual values of log~N(LI) among the M stars, all on the upper locus,
range from 2.19 to 3.08 (in NLTE), with a mean of 2.68 $\pm$0.28 
(3.26 $\pm$0.15 in LTE), where the uncertainties are the standard deviation 
of the means. The uncertainly on any single log~N(Li) value is about 0.2~dex.
The mean Li abundance is consistent with the expected age of the
cloud complex ($\leq$3Myr).
The wide range in log~N(Li) among the individual stars could be due to 
the difficulty of determining the continuum level.
The uncertainty in the adopted T$_{eff}$ of up to 200~K compounds the
uncertainty. 
For KWW~1892 and KWW~975, the  W$_\lambda$(Li) for two different nights
vary by 0.14 and 0.16\AA, which caused 0.38 and 0.52 dex differences 
in the Li abundances.  

The F -- G stars on the lower locus also show high Li abundances.
Since F -- G stars do not burn lithium as fast as late K or M stars do, 
such stars need not be pre-main sequence, but could be 50 -- 100 Myr old 
main sequence stars.

\subsection{Kinematics of PMS stars and CGs}
Radial velocities were derived using Li~I $\lambda$6707.8 \AA\ and 
Ca~I $\lambda$6717.685 \AA\ lines.  The FWHM of a fiber of the HYDRA/echelle 
spectrograph is about 6 pixels (0.96\AA, $\sim$43.9~\kmsec), however we can 
derive radial velocity good to about 3 \kmsec\ by centroiding.
To estimate radial velocities, we derived $\delta \lambda$ ($\lambda_{obs}$ 
$-$ $\lambda_0$), where $\lambda_{obs}$ is observed wavelength, and
$\lambda_0$ for Li~I and Ca~I are from Rao \& Lambert (1993).
The heliocentric correction of $\sim$-8.2~\kmsec\ was then applied. 
To compare the velocity with V$_{LSR}$ (velocities with respect to 
the Local Standard of Rest) for the CGs, we then derived V$_{LSR}$ 
for all the stars.  The correction to V$_{LSR}$ is about -17.3 \kmsec\
\citep{lan80}.  We present the radial velocities in Table~\ref{tbl:EW}.

The V$_{LSR}$ of the K -- M type PMS stars range from 4.2 -- 8.9 \kmsec, 
with a mean measurement error 2.9 \kmsec. This is similar to that of 
the CG clouds \citep{nie98} of 5.1 -- 7.4~\kmsec. 
This strongly supports the association of the PMS stars with the CG~30/31/38
clouds.  The 2.7 \kmsec V$_{LSR}$ of XRS~10 is not significantly lower than
the rest of the PMS  stars.  The radial velocities of the 
stars along the lower locus are inconsistent with $V_{LSR}$ of CGs, 
unless some of these are single-lined spectroscopic binaries.  

\subsection{Spectroscopic Binaries}
Binary stars yield fundamental astrophysical parameters: stellar masses
and occasionally, radii.  It is very difficult to estimate masses and ages 
of PMS stars by only fitting models, as it is the case in our study.
Dynamical determination of masses of binary PMS stars would enable us to
better estimate the ages. 

Spectroscopic binaries have short (few days) periods. This means that the 
separation of two stars are very small. \citet{may84} suggested that 
the dynamical evolution of the short period low mass binary systems depends 
on their PMS history.  These spectroscopic binaries are, therefore, a very 
important resource to help us better understand the star formation history 
of multiple systems.

We identified two double-lined spectroscopic binary systems (SB2): 
KWW~1637 and KWW~1125 (in Figures~\ref{allspectra-b}, \ref{allspectra-d}).
The separation between the Li~I 6708 \AA\ lines is $\sim$0.7 -- 0.8 \AA\ 
($\sim$33 \kmsec) for XRS~6, and $\sim$4.5 \AA\ ($\sim$200 \kmsec) for KWW~1125.
XRS~6 is a strong X-ray source which also shows significant X-ray variability.
KWW~1125 has broad \Ha and Li absorption lines with a large separation between
the lines of the two components, which means that this system may have a 
very short, $\sim$few days, orbital period.  

HBC~554, also known as P\Ha14 \citep{pet87}, is a visual binary star 
\citep{rei93}.  Therefore, of the 17 stars listed in the Table~\ref{tbl:EW}, 
there are at least three binary systems.  
Unfortunately, we have only single epoch observations on these systems;  
we can not determine any orbital parameters with only the current data.

\subsection{Color-Color Diagrams and the Circumstellar Disk Fraction}
In the color-color diagrams (Figures~\ref{ccd1}, \ref{ccd2}) 
we plot the young stars using filled circles (K -- M stars from the upper
locus) and triangles (lower locus).  In Figure~\ref{ccd1} the X-ray 
sources are plotted as crosses and asterisks for comparison.  

Among the stars in the upper locus, only KWW~873 shows a convincing 
near-IR excess in both Figures~\ref{ccd1} and \ref{ccd2}, 
with H -- K~$\sim$~0.2~mag.  
KWW~1892 and KWW~1637 may have marginal excesses.  KWW~1892 is clearly 
a CTTS, based on its \Ha equivalent width.  However, a small excess 
(H -- K $\lesssim$0.1) of KWW~1892 is within the uncertainties of spectral 
type and A$_V$ measurements.  KWW~1637, a SB2 binary system, 
shows variability in X-ray, and also between the near-IR photometry 
from this study and 2MASS photometry, therefore it is unclear if this 
system has near-IR excess (H -- K $\sim$0.1) due to a circumstellar disk.  
Four upper locus stars lie along the CTTS star locus \citep{mey97} 
in Figure~\ref{ccd2}, but they have reddened photospheres with
A$_V$ values between 0.5 -- 1.3, and do not have near-IR excess. 
The stars in the lower locus (triangles) have normal photospheric
colors within the scatter.  

A near-IR excess may indicate the presence of an inner circumstellar disk
around a young star, \eg, \citep[and references therein]{hil03}.  
In our color-color diagrams, one PMS star among the eleven on the upper 
locus shows a convinving near-IR excess.  
The circumstellar disk fraction for this group of stars is 
$\sim$9\%. The disk fraction derived using only JHK colors will under-estimate
the true disk fraction, especially among the cooler stars.
therefore we consider the $\sim$~9\% as a lower limit. 
This low circumstellar disk fraction is consistent with those seen in 
star forming regions where O stars are present, such as 
Upper-Sco \citep{wal94,hil03} or NGC~2362, which shows about a 10\% 
disk fraction at the age of $\sim$~5~$\pm$1 Myr. 
However it is lower than some other known star forming regions, such as 
NGC~2264 ($\sim$55\% at $\sim$~3 Myr, Haisch et al. 2001), and the
TW Hydrae association ($\sim$~20\% at $\sim$~10~Myr). 

The survival time of inner circumstellar disks in T associations is suggested 
to be about 3 -- 10~Myr, however exceptions exist in some cases which show 
very low circumstellar disk fraction in young ($<$3~Myr) associations, \eg, 
\citep[and references therein]{wal88,hai01,hil03}. 
When exposed to strong UV radiation from nearby O stars, surrounding
reservoir gas that supplies gas to the circumstellar disks evaporate
quickly.  By losing reservoir gas due to the photoevaporation, 
those circumstellar disks exposed to UV radiation would likely have 
shorter lifetime than the systems without a source of
UV radiation nearby.  
\section{Discussion}  
We have identified  16 young low mass stars along the line of sight
to the CG~30/31/38 complex. These stars trace out two distinct loci in the
CMD.

The eleven stars that lie along the upper locus have K -- M spectral types.
Their Li abundances are consistent with an age $\lesssim$~5~Myr.
Their location in the CMD is consistent with a 5~Myr isochrone at the
$\sim$200~pc distance of the GC complex.
The 4.2 to 8.9~\kmsec\ ($\pm$2.9~\kmsec) stellar radial velocities agree well
with the 5 -- 7.5 \kmsec\ radial velocities of the CG~30/31/38 complex
\citep{sri92,nie98}.  
This supports the interpretation that these eleven PMS stars are physically
associated with the CG complex.

The stars on the lower locus must be in the background, whether or not
they form a coherent group.  However, they show neither coherent radial 
velocities (see Table~\ref{tbl:EW}) nor a spatial concentration.
We suggest that these stars are unlikely to be associated with CGs or with
the 11 PMS stars along the upper locus.  These stars all have spectral 
types earlier than K and all lack \Ha emission. Five of the F -- G stars 
exhibit strong Li~I absorption lines.  Of the three X-ray sources on the
lower locus (KWW~1055, XRS~11, and XRS~12) only KWW~1055 shows
Li~I $\lambda$~6708 \AA\ line.  

KWW~314, which lies on the lower locus, is an early-type star.
We obtained a low dispersion (4.3/AA/ resolution) blue spectrum using
the SMARTS 1.5m RC spectrograph in February 2004. The spectral type is A3; 
the extinction E$_{B-V}$ is 0.9 $\pm$0.2~mag. There is no emission above 
the continuum between 3500 and 5100\AA.
The $(B - V) = 1.11$ is consistent with a reddened early~A star reddened by 
$A_V \sim$~3~mag.  This suggests an intrinsic $V \sim$~12.
The distance modulus then places this star at $\sim$1.2~kpc.
The extinctions to the other stars on the lower locus are much smaller, 
and inconsistent with that of KWW~314.
We conclude that the lower locus is not a physical association, but a
spurious alignment of 6 young, possibly PMS, background stars and a
perhaps unrelated A3e star.

The log($f_X/f_{bol}$) values of the PMS stars imply that all the K and 
M stars detected in X-rays are active PMS stars. Since the X-ray detections 
are only slightly above the sensitivity limit, we are likely seeing
only the peak of the luminosity distribution if the log($f_X/f_{bol}$) 
relation of this region is similar to that of other star forming regions, 
\eg, \citet{fla02,fla03}.
If we assume that the log($f_X/f_{bol}$) of this star forming region follows
the distribution of Orion Nebula Cluster (ONC) \citep{fla02,fla03}, in the
mass range of 0.2\sm\ -- 1.0\sm, there should be roughly six times as many
PMS stars, all with log($f_X/f_{bol}$)~$<$~-3.0, as we have detected here.
This suggests that deeper X-ray exposures may find about 35 more
PMS stars with log($f_X/f_{bol}$) $<$ -3.0 in this region.

The $V, V-I$ CMD (Figure~\ref{cmd_spec}) shows a cut-off of the upper locus 
at $V~\sim$~18, below which no stars are found. The optical photometry is
complete to $V~\sim$~20, therefore the lack of stars below $V~\sim$~18
(M~$\sim$~0.2~\sm) is significant. This may be due to selective reddening of
lower mass stars, or it may be interpreted as an evidence of a truncated 
mass function below a certain mass.  
Deeper and more complete X-ray and IR observations 
will help us to obtain more complete census of PMS stars to answer whether
or not the mass function of this star forming region is indeed truncated,
and differ from other star forming regions.

In the color-color diagrams (Figures \ref{ccd1}, \ref{ccd2}), 
about 9\% of the PMS stars show the near-IR excesses indicative of 
an inner circumstellar disk. This low circumstellar disk fraction,
also seen in other OB associations, may be due to the influence of 
ionizing sources, such as \zpup, the progenitor of Vela SNR,
and other O stars.  The parent gas cloud, which feeds the accretion disks, 
may be destroyed by the photoevaporation process from nearby O stars
at the early stages of PMS evolution.  This photoevaporation could result in 
a short disk survival time as well as low final masses of those stars.
The lifetime of truncated disks with a mass of 1~\sm\ is about 1~Myr  
if the mass accretion rate onto the star is about $10^{-6}$~\sm~yr$^{-1}$.
Therefore, about 1~Myr after the disk ceases to accrete from the surrounding
cloud the circumstellar disks should dissipate, which may explain the low 
circumstellar disk fraction seen here compared to that of quiet star forming
regions like Taurus. 

Three PMS stars show evidence of accretion in their \Ha emission line profiles.  
Even though their near-IR excesses are not significant, 
they may still possess circumstellar disks.
We suggest that the circumstellar disk fraction and disk lifetime depends
not only on age, but also the environment.  Mid-IR photometry
will help us to derive a more reliable estimate of the
circumstellar disk fraction of these systems.

We identified two double-lined spectroscopic PMS binary systems (KWW~1637 and 
KWW~1125).  KWW~1953 was observed twice on different nights:  the 
Li~I $\lambda$6708 line profiles are very wide (about 2\AA) and are also varying.  
This may due to the fast rotation of the star, or could also be a potential
double-lined binary that were not completely resolved. 
\subsection{Star formation in photoevaporating globules \& Star formation
            history of CG~30/31/38 complex}
For globule-size triggered star formation, the triggering mechanism can be 
shock compression by a supernova explosion \citep{van96}, an expanding H~II 
region \citep{fuk00}, radiation driven implosion \citep{lef94,lef95,lef97}, 
and the stellar wind from massive stars \citep[and references therein]{bos95}.  
Ionizing photons from OB stars photoevaporate the molecular clouds, 
destroying giant molecular clouds. However before the total destruction of 
whole cloud occurs, pre-existing dense parts or cores of the parent molecular 
cloud can survive and form one star or a group of stars.  

The CG formation scenario was discussed by \citet{ber90}.  \citet{lef94}
imply that these small globules in H~II regions are evaporating and 
being compressed due to the external ionizing sources.  
Theoretical studies involving ``rocket effects'' and the ``radiation driven 
implosion'' models \citep{oort55, rei83, lef94} 
explain how a shock can drive into a spherical clump, compress it, and produce 
the cometary shape of clouds in pressure equilibrium with surrounding 
photoevaporating gas. The globule then can be accelerated away from the 
UV source as it photoevaporates.

Whether or not small scale triggered star formation can happen in globules 
depends on a number of physical conditions, such as the time scale needed to 
form a star, the rate of photoevaporation, the shock velocity, magnetic field 
strength and orientation, etc. The theoretical studies 
mentioned above show that star formation can be triggered in globules on a
time scale short compared to the time needed to destroy or evaporate 
globules via UV radiation and/or shocks.   

Here we suggest a simple star formation history of the CG~30/31/38 complex 
in the broader context of the star formation history of the Gum Nebula.

About 6 -- 15 Myr ago, young clusters (now Vela OB2 and Tr 10) formed
in a giant molecular cloud; OB stars in these clusters ionized their 
neighboring clouds and the H~II region expanded (the Gum Nebula).  
Their wind-blown shells may have been comparable in size to those commonly 
seen in the giant H~II regions of other galaxies, such as those seen 
in the {\it HST} image of NGC 4214.  

Approximately 5~Myr ago, the clouds photoevaporated in the UV radiation field
of the luminous O stars, while low mass
stars formed in dense cores of the pre-CG~30/31/38 complex.
Under the strong UV radiation from the O stars, the surrounding gas reservoir
of these young stellar objects began to dissipate.  This resulted in 
accretion disk lifetime shorter than those of stars
forming far from sources of ionizing radiation. 

Some 1 million years ago the runaway early O star \zpup\ moved closer to 
this CG complex, dissipating the clouds surrounding of these young stars.
The kink in the direction of the CG~31 tails is consistent with direction 
of motion of \zpup.  By now, the UV radiation has evaporated most of 
the clouds, though a few dense cores remain. The most recent episode of 
triggered star formation has occurred in the dense core of CG~30, as
evidenced by the Herbig-Haro object HH~120.  Note that the average age
of this kind of HH objects are suggested to be $\lesssim$10$^5$ years,
e.g., \citep{rei00}.  

\section{Summary}

1. Using multi-wavelength photometry and spectroscopy, we identify
   14 new PMS stars, adding to the three previously known \Ha-emitting
   stars in the  CG~30/31/38 region.   The spectroscopic sample is 
   complete to V$\sim$16.5; the optical limiting magnitude is about 20.

2. The stars in the CG~30/31/38 complex lie along two distinct loci in the
   CMD.

3. The upper locus has an age of $\lesssim$ 5 Myr at d = 200~pc.
   The spectral types of the stars on the upper locus are K6 -- M4.  
   The mean Li abundance for M stars, logN(Li) = 2.68 $\pm$0.28
   for NLTE and 3.26 $\pm$0.15 for LTE),
   implies an age of 2 -- 5 Myr and supports the youth of the PMS stars.
   The radial velocities of these stars are consistent with the 
   published velocities of CG 30/31/38 cloud complex. 

4. The lower locus has an age of $<$ 100~Myr at d$ \sim 2$~kpc.
   The stars along the lower locus have spectral types F -- G, with one
   Ae star.
   With an exception of KWW~1055, these stars have radial velocities that
   are not consistent with those of the CG 30/31/38 complex.

5. The $\sim$9\% circumstellar disk fraction derived using optical and
   near-IR excesses is consistent with those seen in other star forming 
   regions with O stars, such as the Upper-Sco, and is lower than those 
   typical of quiet regions, such as Taurus. This suggests that the loss 
   of the gas reservoir for accretion disks at this
   early stage of star formation is due to the radiation from the OB stars.

6. We find two double-lined spectroscopic binaries in the sample.  
   Including the known visual binary HBC~554, the observed multiplicity fraction
   among the 17 PMS stars is 18\%, which sets the lower limit
   to the true multiplicity fraction in this region.

8. We conclude that there have been at least two episodes of star formation 
   in the CG~30/31/38 region:
   (1) On-going star formation triggered by UV radiation from OB stars 
       in the head of CG~30 cloud, as exemplified by HH~120.
   (2) $<$ 5~Myr old low mass (0.2 -- 1 \sm) PMS stars that outline the
       CG~31 complex, whose formation may have been triggered by pre-existing 
       O stars like the progenitor of Vela SNR and $\zeta$~Pup, 
       as well as other OB stars in Vela OB2 and Tr 10.
%
%
\acknowledgments
{
We thank Dr. William Sherry at NSO/NOAO and Dr. Rob Jeffries 
at Keele University for helpful discussions.  We also thank CTIO/4m crew
who helped us to use HYDRA multi-fiber spectrograph. 
This research was supported in part by NASA grant
NAG5-1594 to SUNY Stony Brook.
}
\appendix
\section{Notes on individual PMS stars}
In this appendix we discuss some of the spectroscopically confirmed PMS stars.
Coordinates, optical, and near-IR photometry are in Tables~\ref{tbl:tablePMS_2}
and ~\ref{tbl:SpecIR_PMS} in the discussion section.

\noindent {\bf KWW~464} \\  
This star 17\arcsec\ away from XRS~1, a 4~$\sigma$ deviation.  
The M3 spectral type is based on the TiO band strengths.
$W_\lambda$(Li~I) $\sim$0.66\AA\ indicates that this is a young star that is 
unlikely to be older than 5 - 10 Myr.
The equivalent width of \Ha is $\sim$-4\AA, therefore it is a NTTS/WTTS
(see Table~\ref{tbl:EW}).

\begin{figure}[t]
\figurenum{5a}
\plotone{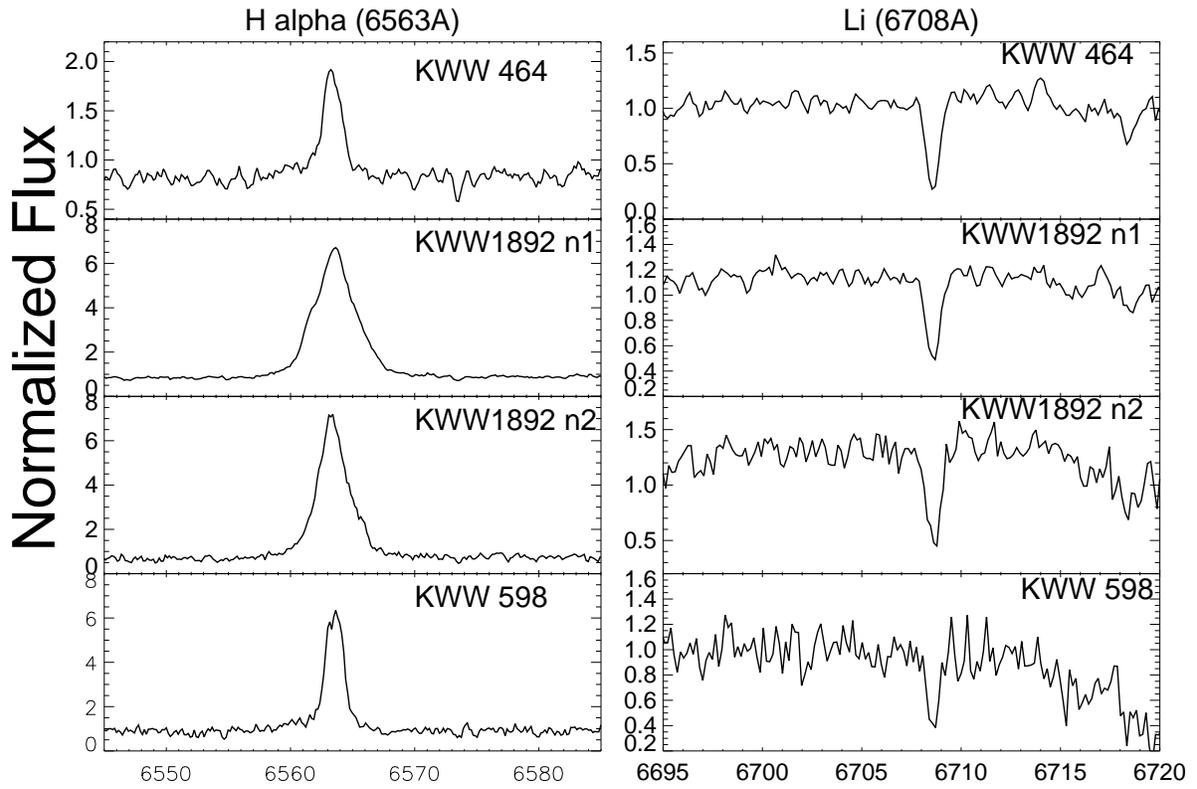}
\caption{HYDRA/echelle spectra of confirmed PMS stars associated with
cometary globules. Continuua are flattened and normalized to unity.
\label{allspectra-a}}
\end{figure}

\noindent {\bf KWW~1892 $=$ XRS~2 $=$ HBC~553} \\
Figure~\ref{allspectra-a} shows two spectra of KWW~1892 (XRS~2; M1 -- M2),
obtained on consecutive nights. This object, also known as HBC~553
\citep{pet87,her88} has the strongest \Ha emission line
($W_{H\alpha} \sim$-30\AA) in our sample. The \Ha emission line profile 
shows changes on the second night, while the absorption line profiles do not.  
The FWHM of \Ha line varied from 4 -- 5\AA\ (night~1) to 3\AA\ (night~2). 
Based on $W_{H\alpha}$, this is the only classical T~Tauri star in our sample.

\noindent {\bf KWW~1637 $=$ XRS~6} \\
Li~I $\lambda$ 6708\AA, Ca~I $\lambda$ 6717\AA, and other Fe~I lines all are 
doubled, strongly indicating that it is a double-lined spectroscopic 
PMS binary system which has 0.8 -- 1\AA\ separation (40 -- 45~\kmsec) at 
this epoch. The X-ray emission was strong and variable.  
 
\noindent {\bf KWW~873 $=$ XRS~7} \\
Although $W_{H\alpha}$ is less than 10\AA, the line profile, with a
blue shifted absorption reversal, implies a strong wind in this system.

\begin{figure}[t]
\figurenum{5b}
\plotone{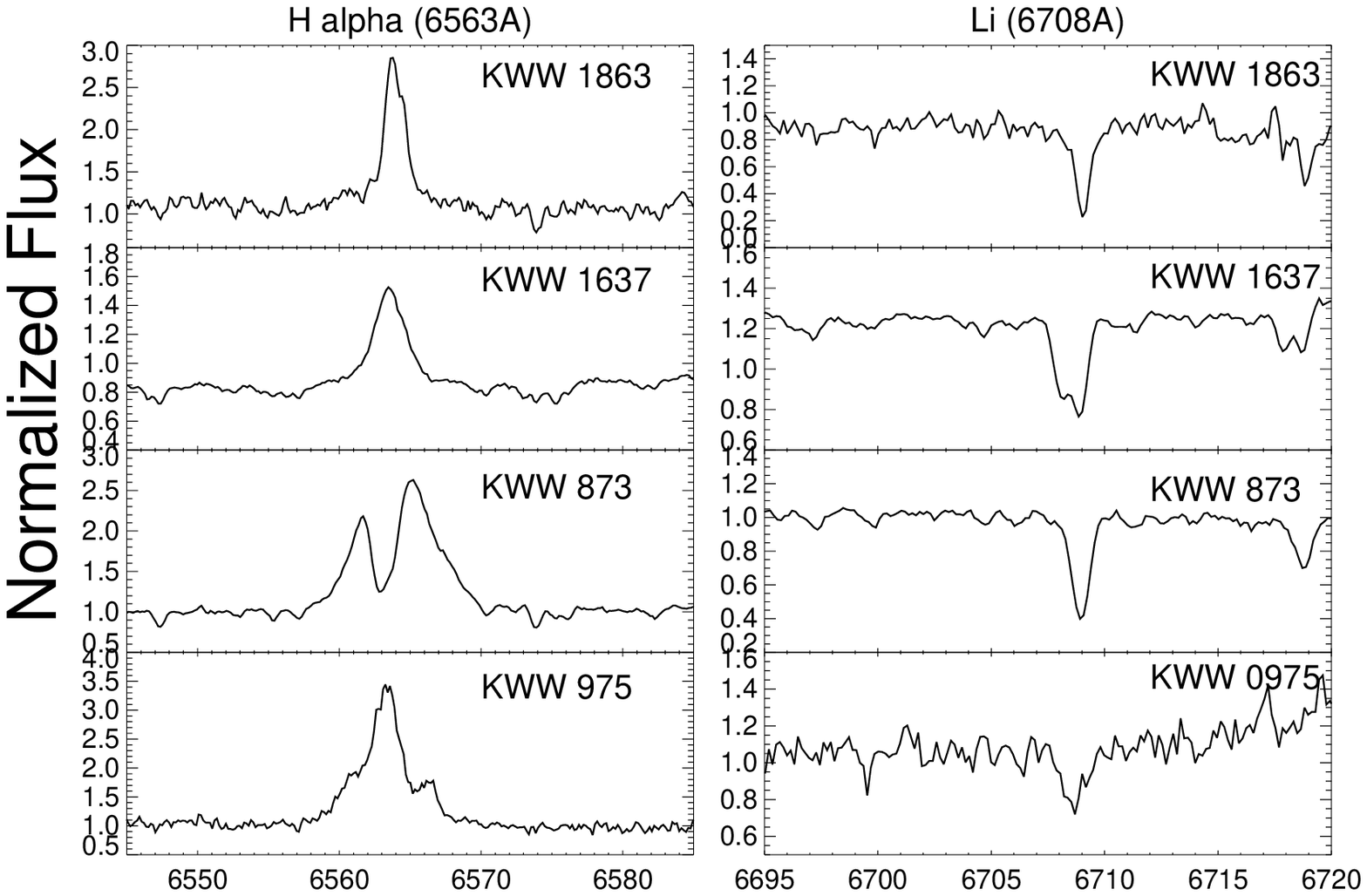}
\caption{HYDRA/echelle spectra of confirmed PMS stars associated with
cometary globules.
\label{allspectra-b}}
\end{figure}

\begin{figure}[t]
\figurenum{5c}
\plotone{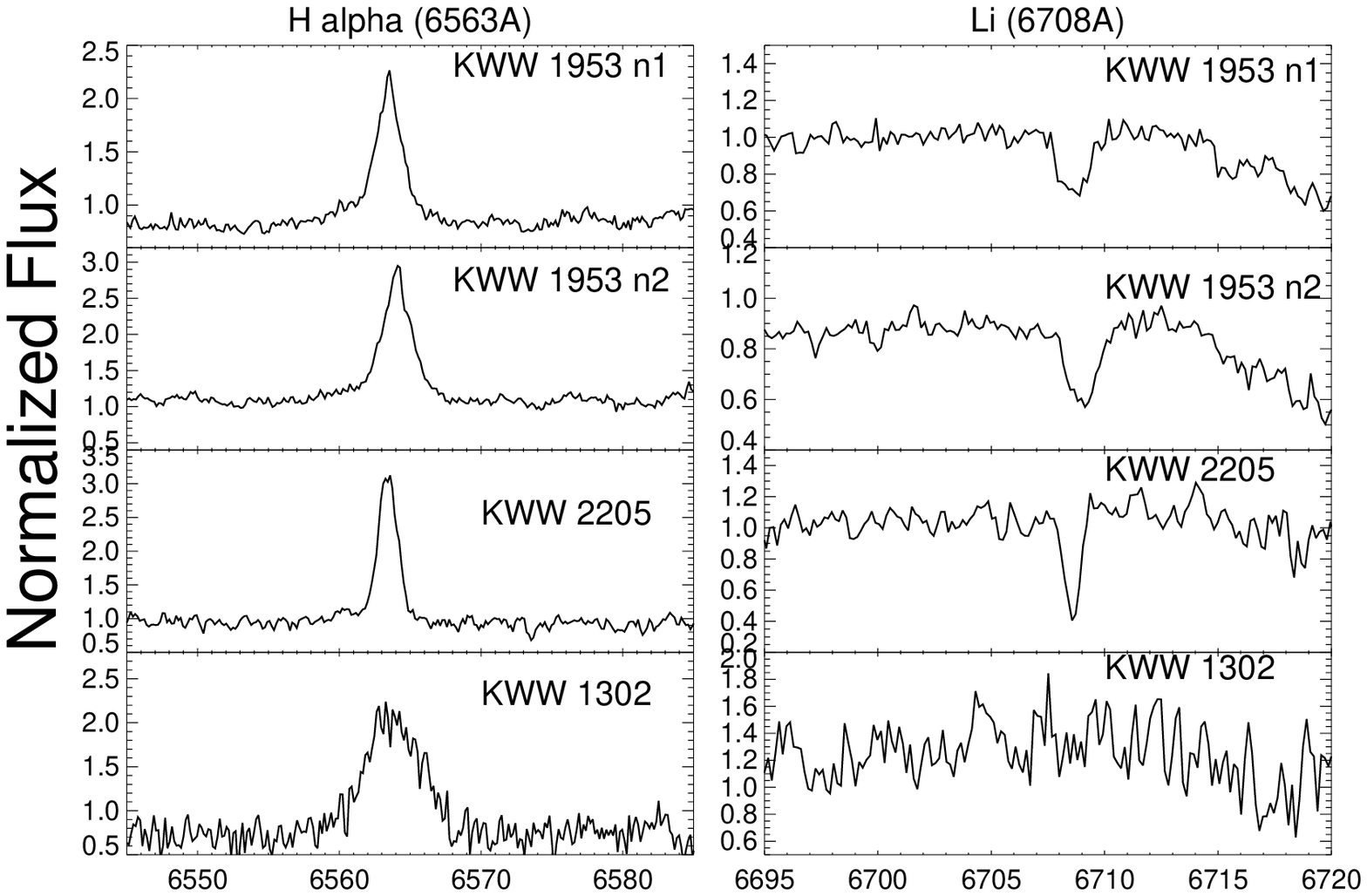}
\caption{HYDRA/echelle spectra of confirmed PMS stars associated with
cometary globules.
\label{allspectra-c}}
\end{figure}

\noindent {\bf KWW~975 $=$ HBC~554}\\
The spectrum of KWW~975 (HBC 554) in Figure~\ref{allspectra-b}  
shows asymmetric \Ha and Li~I line profiles.  The \Ha line profile has dips 
(stronger on the redshifted side), which could indicate accretion or 
infalling gas in the system.  Background sky emission subtraction affects only 
the narrower component, and not the broader line profile.  \citet{her88} 
identified HBC 554 as a visual binary system, and classified the spectral 
type of this star as M~1.5, which is consistent with our spectral type 
estimate (M~2).  

\noindent {\bf KWW~1953} \\
The Li~I 6708\AA\ line and Fe~I lines show asymmetric and broad line profiles.
KWW~1953 may be a rapidly rotating star, or a unresolved double line binary 
system.

\begin{figure}[t]
\figurenum{5d}
\plotone{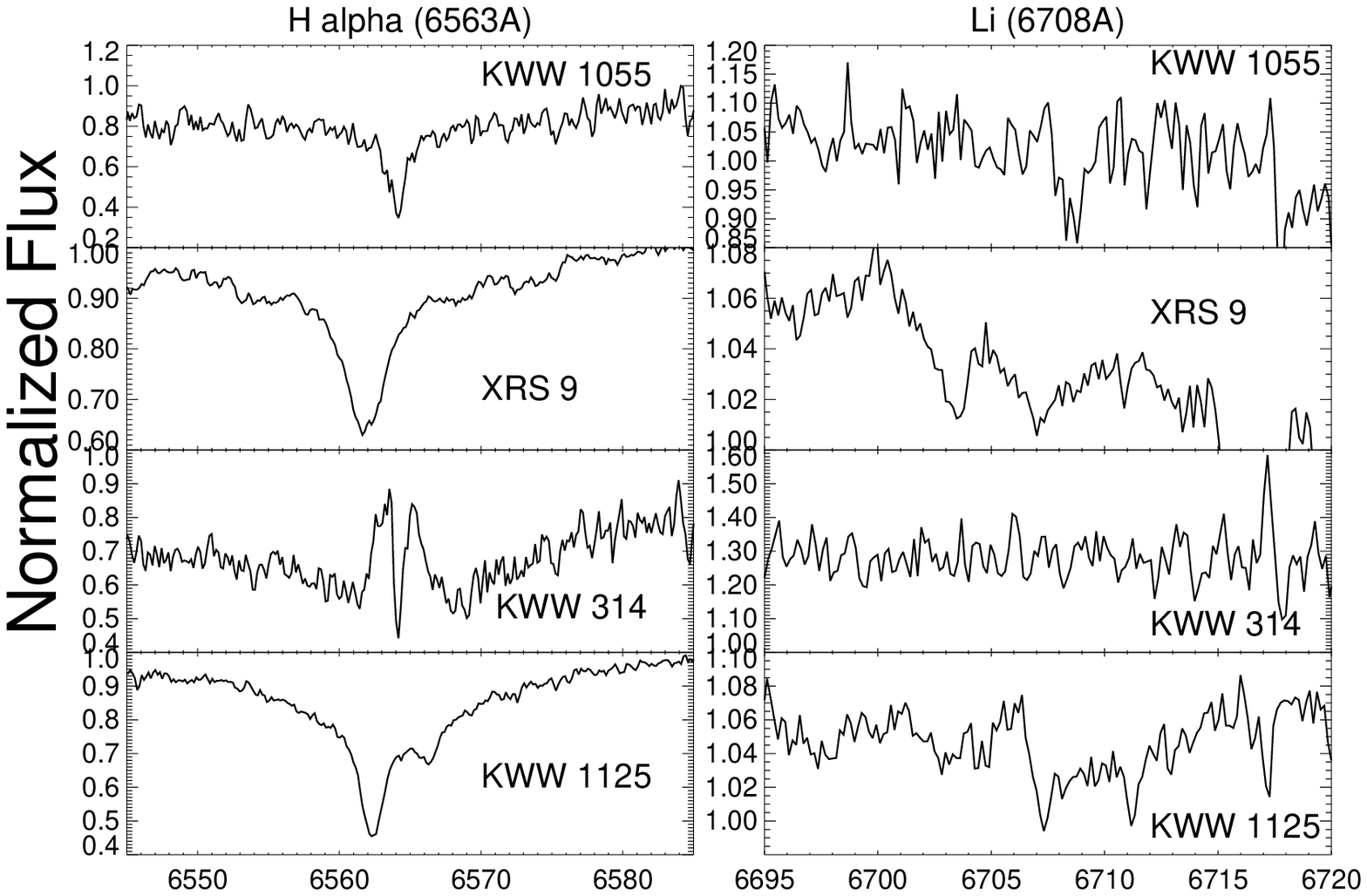}
\caption{HYDRA/echelle spectra of young stars that are unlikely to be
associated with the CG 30/31/38 complex.
\label{allspectra-d}}
\end{figure}

\begin{figure}[t]
\figurenum{5e}
\plotone{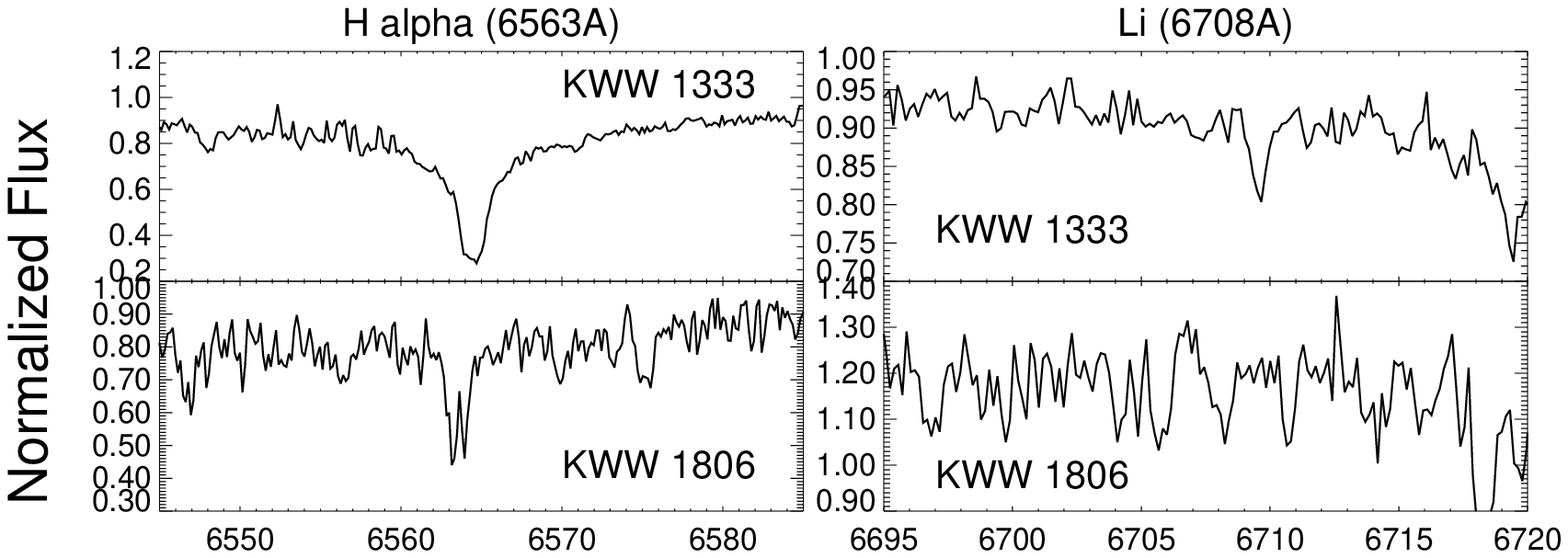}
\caption{HYDRA/echelle spectra of young stars that are unlikely to be
associated with the CG 30/31/38 complex.
\label{allspectra-e}}  
\end{figure}

\begin{figure}[t]
\figurenum{6}
\plotone{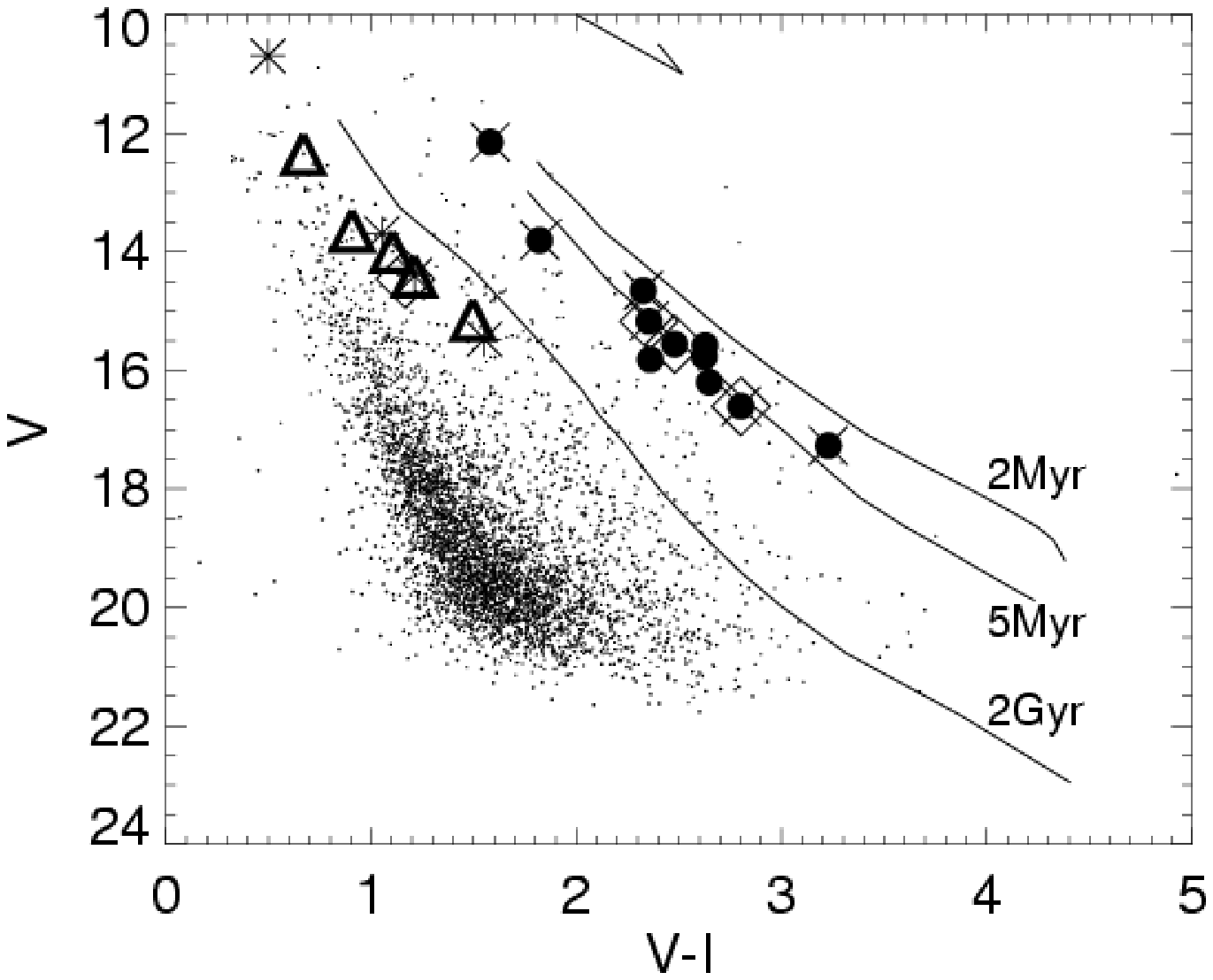}
\caption{  
Spectroscopically confirmed PMS stars in the $V, V-I$~CMD.
This is basically the same plot as Figure~2. Filled circles and
open triangles are spectroscopically confirmed PMS stars including 
KWW~1043 (HBC~555) along the upper PMS locus, and the A3e star (KWW~314) 
along the lower locus.  The stars in the upper PMS locus (filled circles) 
are mainly K and M stars, and appear under the head of CG~31.
The stars in the lower locus (open triangles) are F -- G type stars, and
appear primarily around the CG~30.  The asterisks are optically-unobserved 
X-ray sources. The isochrones are the same as in Figure~\ref{cmd_phot}.
\label{cmd_spec}}
\end{figure}

\begin{figure}[t]
\figurenum{7}
\plotone{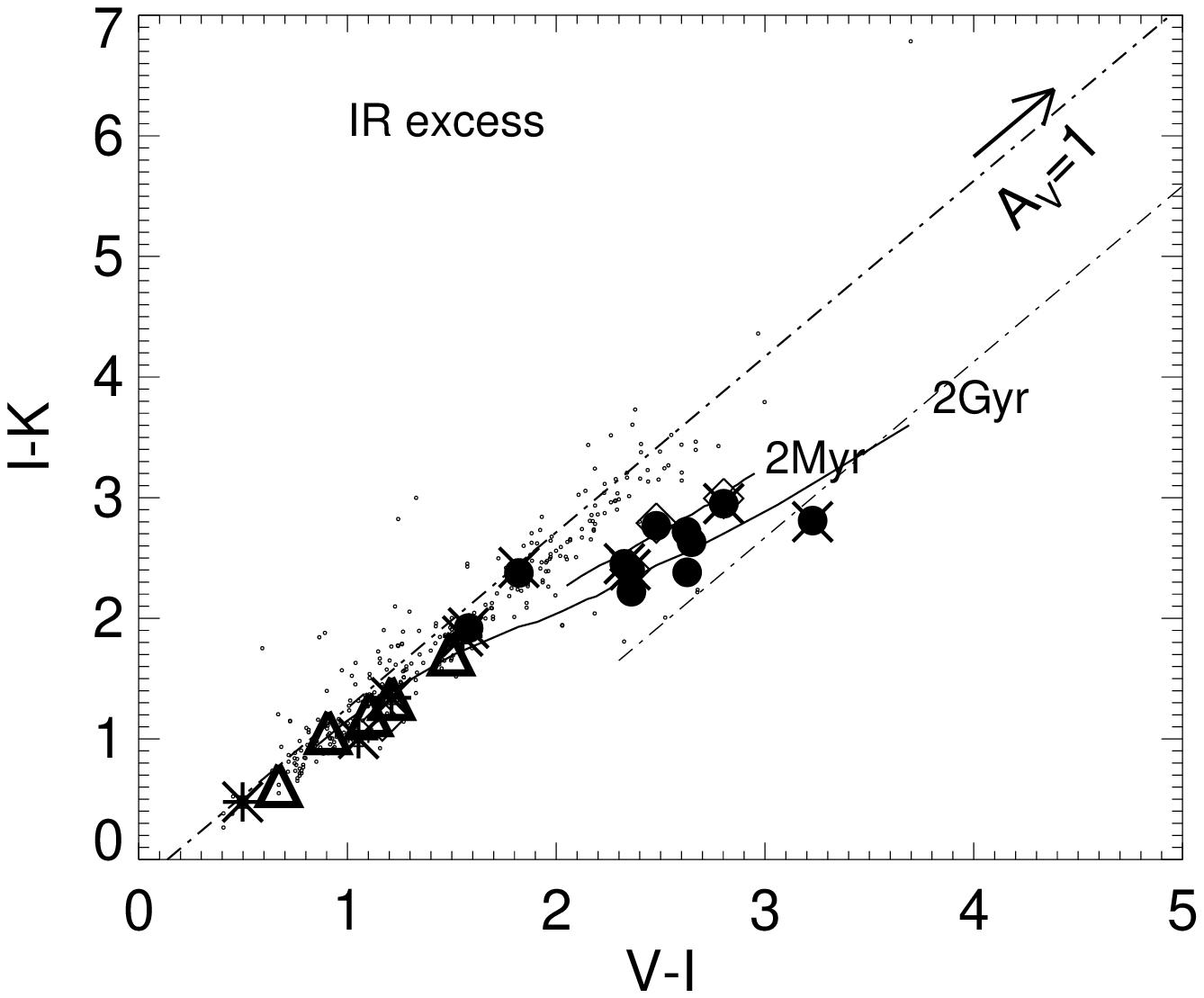}
\caption{
The $I-K, V-I$ color-color diagram.  
Filled circles mark spectroscopically-confirmed PMS stars that are 
kinematically associated with the CGs.  The open triangles mark stars 
that lie along the lower locus, including the A3e star KWW~314, and
open diamond symbols are \Ha\ sources.  The isochrones (Baraffe \etal 1998, 
Luhman 1999) are plotted for 2~Myr and 2~Gyr and distance of 200~pc.  
The reddening vector is shown upper right hand side of the panel
for A$_V$ = 1, and the dash-dot lines indicate reddening directions. 
The label ``IR excess'' indicates the region of the plot occupied by 
IR excess objects.  One source KWW~873 shows I -- K excess. 
\label{ccd1}}
\end{figure}

\begin{figure}[t]
\figurenum{8}
\plotone{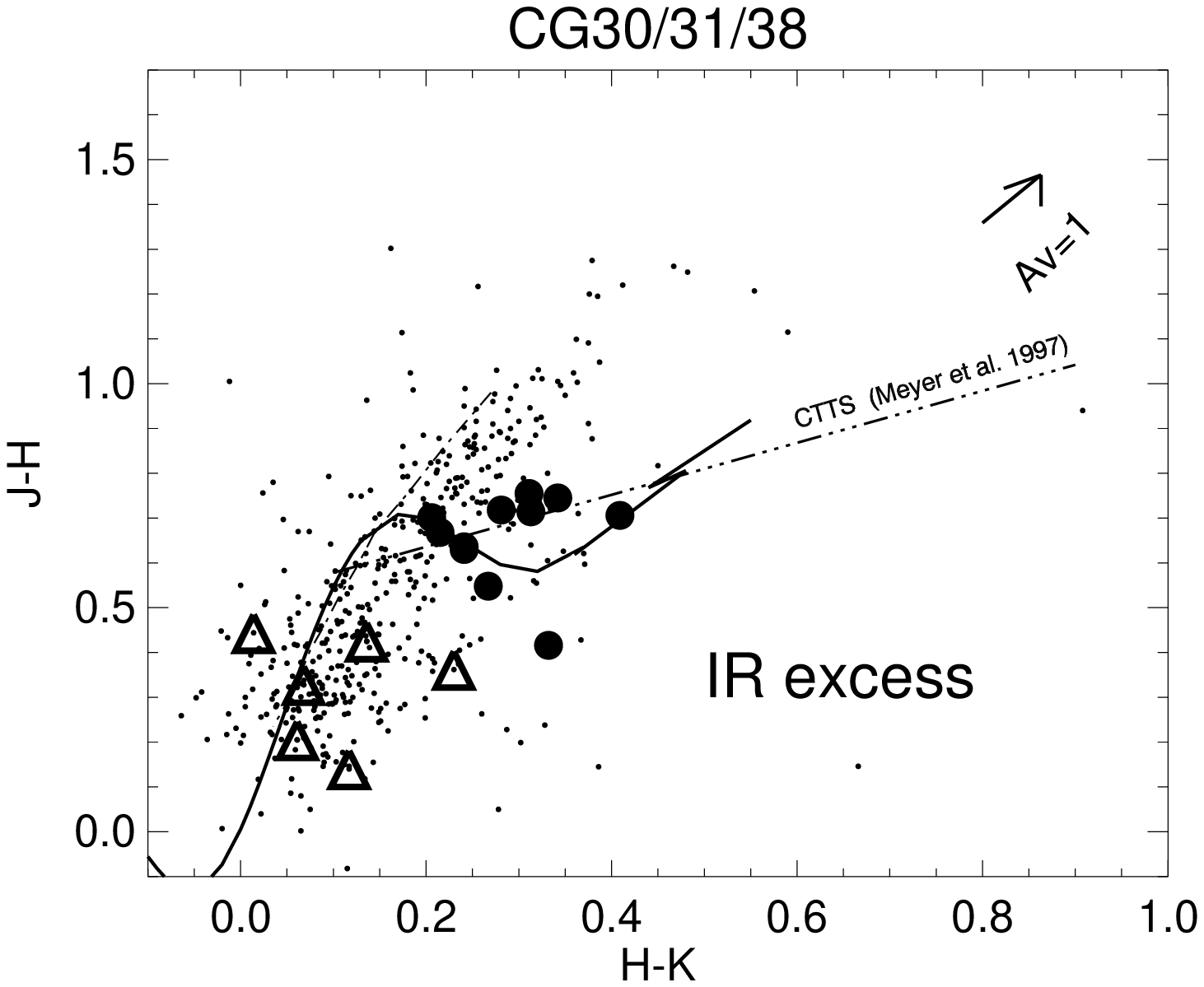}
\caption{
The $J-H, H-K$ color-color diagram of the field. The symbols used
are the same as in Figure~7.  The dash-dot lines are the CTTs locus
from Meyer et al. 1997, and the solid lines indicate approximate 
locations of unreddened dwarf stars using color tables by \citep{ken95}
modified using \citep{luh99, leg01} as in Table~3 in \citep{sher04}.
\label{ccd2}}
\end{figure}

\noindent {\bf KWW~1055 $=$ XRS~10}  \\
The \Ha line is in absorption (Figure~\ref{allspectra-d}).  
The Li line equivalent width is consistent with no Li depletion.
Although appear on the lower locus, and hence at a greater distance,
the radial velocity is marginally consistent with that of the CG 30/31/38
complex.

\noindent {\bf XRS~9} \\
XRS~9 is a fast rotator. It is not a \Ha emission source, but shows a marginal
Li~I line (Figure~\ref{allspectra-d}).  
All the absorption lines are broadened, most likely due to rotation.
All the lines, including the metal lines Fe~I and Ca~I, show flat bottoms 
which may suggest a spot on its surface. 
The radial velocity is inconsistent with those of the
of CGs and associated PMS stars.  It is likely a young star in the
background.

\noindent {\bf KWW~314} \\  
The double-peaked \Ha emission arising from a broad 
\Ha absorption line is typical of a Ae/Be star (and not necessarily of a 
young object).  A low dispersion spectrum confirms that this is an 
A3 ($\pm$2) star with E$_{B-V}$ = 0.9$\pm$0.2~mag. A$_v$ is $\sim$3 mag, 
(see tables~\ref{tbl:tablePMS_2} and \ref{tbl:SpecIR_PMS}), and thus 
$V \sim$12mag.  The distance modulus is then $\sim 10.5$ magnitude 
($\sim$1.2~kpc).  The radial velocity ($V_{LSR}$ = -15.11 \kmsec) is 
inconsistent with that of the CGs and PMS stars.  This is therefore likely 
a background star unrelated to the CG~30/31/38 complex.

\noindent {\bf KWW~1125} \\
This is a young double-lined spectroscopic binary system with undepleted Li
abundances.  The spectrum (Figure~\ref{allspectra-d}) shows broad and double 
peaked \Ha and Li~I absorption lines. Fe~I and Ca~I lines also seem to have 
double lines.  The two components are $\sim$4\AA\ (180 \kmsec) apart 
at the epoch of observation; this is likely to be a short period system 
with a period of a few days. The redshifted component has $\sim$50\% of 
the line strength of the blueshifted component in \Ha, but the Li~I strengths
of both components are similar. 
At this epoch the radial velocities are -65.6 \kmsec\ for 
the blue shifted component and 152.6 \kmsec\ for the redshifted component.
Since only one epoch was observed, we have no information on the
systemic velocity or the mass ratio. 



%
\newpage
\clearpage
\begin{deluxetable}{ccccccrrccrr}
\tabletypesize{\scriptsize}
\tablewidth{0pt}
\tablecolumns{12}
\tablenum{1}
\setlength{\tabcolsep}{0.06in}
\tablecaption{ROSAT/HRI sources in CG~30/31/38 complex \label{Table_1}}
\tablecomments{ Values are taken from
the Standard Analysis Software System (SASS) output.
Note that the uncertainty on the count rate does not directly translate
to a Signal to Noise (S/N) ratio: the former corresponds to a larger
radius containing 90\% of the encircled energy. The significance
of the count rate of the weak sources is decreased because background noise
dominates.  
}
\tablehead{
\colhead{XRS \#} & \colhead{$\alpha$ (2000.0)}      & \colhead{$\pm$} &
\colhead{$\delta$ (2000.0)}    &  \colhead{$\pm$}   &
\colhead{$\Delta$$\alpha$\tablenotemark{a}} &    
\colhead{$\Delta$$\delta$\tablenotemark{b}} &    
\colhead{DC\tablenotemark{c}}  &                 
\colhead{CR\tablenotemark{d}}  &                 
\colhead{$\pm$}                &                 
\colhead{net\tablenotemark{e}} &                 
\colhead{SNR\tablenotemark{f}}                   
\\    
\colhead{ }   &   \colhead{h : m : s}          &  \colhead{s}    & 
\colhead{${\circ}$ : $\arcmin$ : $\arcsec$}    &  \colhead{$\arcsec$} & 
\colhead{s}   &  \colhead{$\arcsec$}    &  \colhead{ }    &
\multicolumn{2}{c}{cts s$^{-1}$}        &  \colhead{cts}  &   \colhead{ }   
}
\startdata
 1 & $08:07:59.8$ & 0.5 & $-35:57:50.5$ & 4.2 & 0.87 & 16.82 &  24 & 0.00756 & 0.00106 & 107 & 3.4  \\ 
 2 & $08:08:21.5$ & 0.5 & $-36:03:37.5$ & 4.2 & 0.65 &  9.58 &  24 & 0.00108 & 0.00051 &  18 & 3.2  \\ 
 3 & $08:08:37.4$ & 0.2 & $-36:09:46.9$ & 1.9 & 0.19 &  3.58 &  12 & 0.00102 & 0.00041 &  17 & 3.6  \\ 
 4 & $08:08:37.8$ & 0.1 & $-36:03:56.8$ & 0.6 & 0.02 &  1.55 &  12 & 0.02042 & 0.00112 & 360 & 15.9 \\
 5 & $08:08:38.8$ & 0.6 & $-36:19:25.1$ & 4.6 & 0.17 &  4.33 &  24 & 0.00138 & 0.00073 &  21 & 3.2  \\
 6 & $08:08:39.0$ & 0.1 & $-36:04:58.5$ & 0.6 & 0.29 &  3.40 &  12 & 0.01937 & 0.00109 & 340 & 15.3 \\ 
 7 & $08:08:45.2$ & 0.2 & $-36:08:38.1$ & 1.3 & 0.20 &  2.14 &  12 & 0.00232 & 0.00046 &  41 & 5.8  \\
 8 & $08:08:46.6$ & 0.6 & $-36:07:41.2$ & 4.5 & 0.22 & 11.56 &  24 & 0.00111 & 0.00038 &  14 & 3.0  \\
 9 & $08:09:02.9$ & 0.2 & $-35:51:28.6$ & 1.2 & 0.02 &  3.12 &  12 & 0.02126 & 0.00149 & 300 & 7.7  \\ 
10 & $08:09:13.1$ & 0.3 & $-36:10:29.0$ & 2.5 & 0.39 &  0.71 &  12 & 0.00085 & 0.00035 &  16 & 3.2  \\
11 & $08:09:22.2$ & 0.2 & $-36:06:47.2$ & 1.9 & 0.05 &  3.67 &  12 & 0.00120 & 0.00038 &  22 & 3.6  \\
12 & $08:09:24.6$ & 0.3 & $-36:13:24.0$ & 2.4 & 0.09 &  0.90 &  12\tablenotemark{g} & 0.00064 & 0.00036 &  
11 & 3.0 \\
13 & $08:09:35.1$ & 0.2 & $-36:13:08.8$ & 1.8 & 0.05 &  2.30 &  12 & 0.00343 & 0.00055 &  60 & 4.7  \\ \hline
\enddata
\tablenotetext{a} {X-ray - optical position offset in RA}  
\tablenotetext{b} {X-ray - optical position offset in DEC}  
\tablenotetext{c} {Detect cell size in arcsec}     
\tablenotetext{d} {Net count rate within the 90\% encircled energy radius}   
\tablenotetext{e} {Net source counts (above background) within the 90\% encircled energy radius}  
\tablenotetext{f} {signal-to-noise ratio within the detect cell}  
\tablenotetext{g} {Not detected in the 24 arcsec detection cell} 
\end{deluxetable}
%
\clearpage
\begin{deluxetable}{ccccccccccccrl}
\tabletypesize{\scriptsize}
\rotate
\tablewidth{0pt}
\tablecolumns{14}
\tablenum{2}
\setlength{\tabcolsep}{0.06in}
\tablecaption{Young stars identified from this spectroscopic study \label{tbl:EW}} 
\tablecomments{ The spectral types and equivalent widths of \Ha $\lambda$6563\AA\ and Li~I
$\lambda$6708\AA\ for PMS stars are presented.
Note that XRS~9, KWW~1055, KWW~1125, KWW~1333, and KWW~1806 are more likely 
50 -- 100 Myr old field stars with strong magnetic activity.
} 
\tablehead{
\colhead{KWW\#}   &   \colhead{XRS\#}   &   \colhead{Sp.}    & 
\colhead{$\alpha$ (2000.0)}      &   \colhead{$\delta$(2000.0)}    &
\colhead{W$_\lambda$(\Ha)}   &   \colhead{W$_\lambda$(Li I)} &  
\colhead{$\pm$\tablenotemark{a}}  &    \colhead{logN(Li)\tablenotemark{b}}    &  
\colhead{A$_V$}  &    \colhead{$\sigma$(A$_V$)}\tablenotemark{c}  &  \colhead{log(\fxfbol)}  &  
\colhead{V$_{LSR}$}  &  \colhead{Note}
\\
\colhead{ }   &   \colhead{ }   &   \colhead{ }    &  
\colhead{ h : m : s }      &   \colhead{ \degr\ : \arcmin\ : \arcsec}    &
\colhead{\AA}   &   \colhead{\AA} &  
\colhead{ }  &    \colhead{(NLTE/LTE)}    &  
\colhead{mag}  &    \colhead{ }  &  \colhead{ }  &  
\colhead{\kmsec}  &  \colhead{ }
}
\startdata
 464  & 1  & M3V  & 08:08:00.667 & -35:57:33.68 &  -2.8 & 0.67  & 0.02 &  2.70/3.31  & 0.2  & 0.2 & 0.0186 & 6.69 &  \\
1892  & 2  & M1V  & 08:08:22.153 & -36:03:47.08 & -26.6 & 0.54  & 0.04 &  2.54/3.08  & 0.3  & 0.3 & 0.0028 & 4.87 & night 1, 
HBC~553\tablenotemark{d} \\
      &        &      &              &          & -31.3 & 0.70  & 0.05 &  3.01/3.35  &  --  &  -- &        &      & night 2 \\
 598  & 3  & M2V  & 08:08:37.586 & -36:09:50.48 & -11.5 & 0.70  & 0.04 &  2.83/3.35  & 0.3  & 0.3 & 0.0147 & 4.21 &   \\
1863  & 4  & M1V  & 08:08:37.824 & -36:03:55.25 &  -2.8 & 0.74  & 0.03 &  3.08/3.41  & 0.4  & 0.2 & 0.0335 & 8.92 &   \\ 
1637  & 6  & K6V  & 08:08:39.286 & -36:05:01.90 &  -2.4 & 0.63  & 0.01 &  3.65/3.41  & 0.1  & 0.1 & 0.0061 & 5.51 & SB2     \\
 873  & 7  & K7V  & 08:08:45.403 & -36:08:40.24 &  -7.9 & 0.68  & 0.02 &  3.57/3.37  & 0.6  & 0.2 & 0.0028 & 4.95 & 
P Cyg profile in \Ha  \\
1043  & 8  & M3V  & 08:08:46.821 & -36:07:52.76 &  --   &    -- & --   &     --      & 0.8  & 0.2 & 0.0019 &  --  &
HBC~555\tablenotemark{d} \\
 975  & -- & M2V  & 08:08:33.870 & -36:08:09.82 & -8.43 & 0.50  & 0.06 &  2.19/3.00  & 1.3  & 0.2 & --     & 6.30 & Visual binary, 
HBC~554\tablenotemark{d} \\
 1302 & -- & M4V  & 08:09:51.778 & -36:23:02.87 & -8.23 &  --   &  --  &     --      & 0.7  & 0.4 & -- &  --   &  \\
 1953 & -- & M3V  & 08:08:26.929 & -36:03:35.45 & -4.24 & 0.56  & 0.03 &  2.39/3.12  & 0.5  & 0.3 & -- &  7.20 & night 1,  \\
      & -- &      &              &              & -4.93 & 0.70  & 0.04 &  2.77/3.35  & --   &  -- &    &  --   & night 2  \\
 2205 & -- & M4V  & 08:08:14.872 & -36:02:09.46 & -4.34 & 0.62  & 0.04 &  2.58/3.23  & 0.3  & 0.3 & -- &  8.03 & \\ 
  --  & 9  & G5V  & 08:09:02.9   & -35:51:28.6  &  3.23 & 0.14  & 0.02 &  2.89/2.86  & --   &  -- & -- & -91.35 & Rapid rotator \\
 1055 & 10 &G2V--K0V& 08:09:13.493 & -36:10:29.71&  2.17 & 0.27 & 0.02 &  3.74/4.00  & 1.0  & 0.2 & 0.0031 & 2.74 &    \\
 314  & -- & A3e  & 08:09:52.001 & -36:00:36.92 &  6.2  &  --   &  --  &     --      & 3    &  -- & -- & -15.11 &  Ae star    \\
      &    &      &              &              & -1.58 &       &      &             & --   &  -- & -- & --     &   \\
 1125 & -- & $<$F8V & 08:09:15.140 & -36:09:14.22 &  5.75 & 0.32 & 0.02 & 3.87/4.00  & 0.12 & $>$0.12 & -- & 152.60 & 
SB2\tablenotemark{e} \\
 1333 & -- & $<$F8V & 08:08:34.749 & -36:06:40.90 &  5.23 & 0.12 & 0.02 & 3.05/3.00  & 0.16 & $>$0.16 & -- &  19.08 &  \\
 1806 & -- & $<$F8V & 08:09:33.525 & -36:13:09.14 &  0.81 & 0.13 & 0.02 & 3.09/3.00  & 0.08 & $>$0.08 & -- &  -10.70 & \\ \hline
\enddata 
\tablenotetext{a}{The errors were determined by measuring \EWLi\ multiple ($>$10) times.}
\tablenotetext{b}{Lithium abundance for both LTE and NLTE model (NLTE from  
Pavlenko \& Magazzu 1996; LTE from Duncan 1991)}
\tablenotetext{c}{The uncertainty in A$_V$ value is mainly introduced by the uncertainty 
                  of spectral type determination ($\sim$0.5 -- 1 subtype for K and M stars, 
                  $>$1 for G type and earlier).  }
\tablenotetext{d}{Pettersson (1987), Herbig \& Bell (1988)}
\tablenotetext{f}{Both components show the same Lithium strength.  The equivalent width here
is for the blueshifted component.}

\end{deluxetable}
\clearpage

\begin{table}
\tablenum{3}
\caption{Optical photometry} \label{tbl:tablePMS_2}
\vspace{0.5cm} 
\singlespace
\scriptsize
\begin{center}
\begin{tabular}{lrrrrrrrrrr} \hline \hline
ID        &$U-B$  &$\sigma (U-B)$  &$B-V$  & $\sigma (B-V)$ & $V$   & $\sigma V$ & $V-R$ &  $\sigma (V-R)$& $R-I$ &$\sigma (R-I)$   \\ \hline
KWW~464   &   -   &   -   &  1.52 &  0.04 & 15.82 &  0.03 &  1.14 &  0.04 &  1.22 &  0.05  \\
KWW~1892  &  1.04 &  0.03 &  1.42 &  0.03 & 15.17 &  0.02 &  1.12 &  0.03 &  1.23 &  0.02  \\
KWW~598   &  0.81 &  0.08 &  1.35 &  0.04 & 17.27 &  0.02 &  1.45 &  0.03 &  1.77 &  0.02  \\
KWW~1863  & -0.29 &  0.47 &  3.01 &  0.34 & 14.65 &  0.01 &  1.10 &  0.02 &  1.22 &  0.03  \\
KWW~1637  &  1.05 &  0.02 &  1.20 &  0.02 & 12.15 &  0.01 &  0.92 &  0.02 &  0.66 &  0.03  \\
KWW~873   &  0.99 &  0.02 &  1.35 &  0.02 & 13.81 &  0.01 &  0.92 &  0.02 &  0.90 &  0.02  \\
KWW~1043  &  0.36 &  0.04 &  1.31 &  0.03 & 16.61 &  0.02 &  1.25 &  0.03 &  1.55 &  0.06  \\
KWW~1055  &  0.41 &  0.02 &  0.96 &  0.01 & 14.40 &  0.01 &  0.58 &  0.02 &  0.63 &  0.03  \\
KWW~975   & -0.10 &  0.03 &  1.15 &  0.02 & 15.56 &  0.01 &  1.12 &  0.02 &  1.35 &  0.03  \\
KWW~314   &   -   &   -   &  1.11 &  0.01 & 15.14 &  0.00 &  0.73 &  0.00 &  0.77 &  0.00  \\
KWW~1806  &  0.71 &  0.02 &  0.94 &  0.01 & 13.99 &  0.01 &  0.68 &  0.22 &  0.43 &  0.06  \\
KWW~1125  & -0.01 &  0.02 &  0.47 &  0.02 & 12.35 &  0.01 &  0.30 &  0.02 &  0.37 &  0.02  \\
KWW~1302  &  -    &   -   &  1.39 &  0.01 & 15.76 &  0.00 &  1.20 &  0.00 &  1.42 &  0.00  \\
KWW~1333  &  0.24 &  0.02 &  0.65 &  0.01 & 13.64 &  0.01 &  0.47 &  0.02 &  0.44 &  0.02  \\
KWW~1953  &  1.05 &  0.03 &  1.37 &  0.03 & 15.58 &  0.02 &  1.21 &  0.02 &  1.42 &  0.02  \\
KWW~2205  &  0.49 &  0.49 &  2.38 &  0.35 & 16.20 &  0.02 &  1.25 &  0.02 &  1.40 &  0.17  \\ \hline 
\end{tabular}
\end{center}
\end{table}

\clearpage
\begin{table} 
\tablenum{4} 
\caption{Near IR photometry} \label{tbl:SpecIR_PMS}
\vspace{0.5cm} 
\singlespace 
\scriptsize 
\begin{center}
\begin{tabular}{lrcrcrcrcrcrc} \hline \hline
ID  & $K$ &$\sigma K$& $J-K$ & $\sigma (J-K)$ &$ H-K$ & $\sigma (H-K)$ & $J^a$ & $\sigma J$  & $H^a$  & $\sigma H$  & $K^a$ & $\sigma K$ \\ \hline
KWW~464  &11.237& 0.040& 0.907& 0.042& 0.206& 0.048 & 12.126 &  0.024 & 11.392 &  0.026 & 11.173 &  0.026 \\
KWW~1892 &10.416& 0.040& 1.027& 0.042& 0.313& 0.048 & 11.402 &  0.023 & 10.663 &  0.023 & 10.323 &  0.021 \\
KWW~598  &11.232& 0.040& 0.815& 0.042& 0.267& 0.048 & 12.150 &  0.024 & 11.519 &  0.023 & 11.281 &  0.023 \\
KWW~1863 & 9.876& 0.040& 0.999& 0.042& 0.281& 0.048 & 10.873 &  0.028 & 10.194 &  0.033 &  9.940 &  0.024 \\ 
KWW~1637 & 8.650& 0.040& 0.877& 0.041& 0.241& 0.048 &  9.529 &  0.023 &  8.880 &  0.022 &  8.708 &  0.024 \\
KWW~873  & 9.609& 0.040& 1.066& 0.042& 0.311& 0.048 & 10.676 &  0.023 &  9.947 &  0.023 &  9.578 &  0.023 \\
KWW~1043 &10.854& 0.040& 1.115& 0.042& 0.409& 0.048 & 11.925 &  0.028 & 11.135 &  0.025 & 10.628 &  0.024 \\
XRS~9    & 9.383& 0.040& 0.396& 0.042& 0.068& 0.048 &  9.783 &  0.021 &  9.464 &  0.022 &  9.346 &  0.023 \\
KWW~1055 &11.850& 0.041& 0.561& 0.043& 0.136& 0.048 & 12.424 &  0.026 & 12.028 &  0.025 & 11.890 &  0.023 \\
KWW~975$^a$  &10.312& 0.025& 1.087& 0.033& 0.342& 0.041 & 11.387 &  0.022 & 10.640 &  0.025 & 10.299 &  0.023 \\
KWW~314$^a$  &11.937& 0.028& 0.592& 0.038& 0.230& 0.039 & 12.546 &  0.024 & 12.155 &  0.023 & 11.931 &  0.026 \\
KWW~1125 &11.055& 0.040& 0.266& 0.042& 0.061& 0.048 & 11.285 &  0.023 & 11.085 &  0.023 & 11.015 &  0.023 \\
KWW~1302$^a$ &10.421& 0.027& 1.080& 0.040& 0.332& 0.040 & 11.500 &  0.032 & 10.761 &  0.036 & 10.429 &  0.023 \\
KWW~1806$^a$ &11.681& 0.028& 0.458& 0.040& 0.014& 0.037 & 12.262 &  0.028 & 11.820 &  0.031 & 11.678 &  0.027 \\
KWW~1333$^a$ &11.671& 0.029& 0.257& 0.042& 0.117& 0.045 & 12.074 &  0.024 & 11.788 &  0.027 & 11.694 &  0.026  \\
KWW~1953 &10.571& 0.040& 0.883& 0.042& 0.215& 0.048 & 11.438 &  0.023 & 10.722 &  0.022 & 10.510 &  0.023 \\
KWW~2205 &10.923& 0.041& 0.871& 0.043& 0.241& 0.048 & 11.780 &  0.023 & 11.122 &  0.022 & 10.840 &  0.021 \\ \hline
\end{tabular}
\end{center}
$a$: $JHK$ data from the 2MASS All-Sky Release Point Source Catalog. 

\end{table}

\newpage

\clearpage

\clearpage

\clearpage

\clearpage
%
\clearpage

\clearpage
\clearpage



\begin{thebibliography}{}
\bibitem[Alves et al.(1997)]{alv97} Alves, J., Hartmann, L., Briceno, C., \& 
        Lada, C. J. 1997, \aj, 113, 1395
\bibitem[Bally et al.(1998)]{bal98} Bally, J., Yu, K. C., Rayner, J. \& Zinnecker, 
        H. 1998, \aj, 116, 1868
\bibitem[Bally et al.(2001)]{bal01} Bally, J., Johnstone, D., Joncas, G., 
        Reipurth, B., \& Mallen-Ornelas, G. 2001, \aj, 122, 1508
\bibitem[Bareffe et al.(1998)]{bar98} Baraffe, I., Chabrier, G., Allard, F., \& 
        Hauschildt, P. H. 1998, \aap, 337, 403
\bibitem[Basri et al.(1991)]{bas91} Basri, G., Martin, E. L., Bertout, C. 1991, \aap, 
        252, 625
\bibitem[Bertoldi \& McKee(1990)]{ber90} Bertoldi, F. M., \& McKee, C. 1990, ApJ, 
        354, 529
\bibitem[Bok \& Reilly(1947)]{bok47} Bok, B. J., \& Reilly, E. F. 1947, \apj, 105, 255
\bibitem[Bok(1977)]{bok77} Bok, B. J. 1977, PASP, 89, 597
\bibitem[Boss(1995)]{bos95} Boss, A.P. 1995, \apj, 439, 224
\bibitem[Brandt et al.(1971)]{bran71} Brandt, J. C., Stecher, T. P., Crawford, D. L., 
        \& Maran, S. P. 1971, \apjl, 163, 99 
1997, in {\it Astrophysical Implications of the Laboratory Study of Presolar Materials}, ed. 
Bernatowicz, T. J., \& Zinner, E. Woodbury, N.Y.: American Institute of Physics, 1997., 665
\bibitem[Chanot \& Sivan(1983)]{cha83} Chanot, A., \& Sivan, J. P. 1983, \aap, 121, 19
\bibitem[Clemens \& Barvainis(1988)]{cle88} Clemens, D. P. \& Barvainis, R. 1988, 
        \apjs, 68, 257
\bibitem[de Zeeuw et al.(1999)]{dez99} de Zeeuw, P. T., Hoogerwerf, R., 
        de Bruijne, J. H. J., Brown, A. G. A., \& Blaauw, A. 1999, \aj, 117, 354
\bibitem[Duncan(1991)]{dun91} Duncan, D. K. 1991, \apj, 373, 250
\bibitem[Elias et al.(1982)]{eli82} Elias, J. H., Frogel, J. A., Matthews, K., Neugebauer, G. 1982, \aj, 87, 1029
\bibitem[Elmegreen(1998)]{elm98} Elmegreen, B. G. 1998, in ASP Conf. Ser. 148, 
        {\it Origins of Galaxies, Stars, Planets and Life}, ed. Woodward, C. E., 
        Thronson, H. A., \& M. Shull (San Francisco: ASP), 150
\bibitem[Feigelson et al.(2002)]{fei02} Feigelson, E. D. et al. 2002, \apj, 574, 258 
\bibitem[Flaccomio et al.(2003a)]{fla02} Flaccomio, E. et al. 2003a, \apj, 582, 398 
\bibitem[Flaccomio et al.(2003b)]{fla03} Flaccomio, E., Micela, G., \& Sciortino, S. 2003b, \aap, 402, 277
\bibitem[Frerking \& Langer(1982)]{fre82} Frerking, M. A., \& Langer, W. D. 1982, 
        \apj, 256, 523
\bibitem[Fukuda et al.(2000)]{fuk00} Fukuda, N., \& Hanawa T. 2000, \apj, 533, 911
        \& Van Buren, D. 2001, PASP, 113, 1326
\bibitem[Gum(1952)]{gum52} Gum, C. S. 1952, Observatory,  72, 151 
\bibitem[Gum(1955)]{gum55} Gum, C. S. 1955, MemRAS, 67, 155 
\bibitem[Haisch(2001)]{hai01}Haisch Jr., K. E., Lada, E. A., \& Lada, C. J. 2001, \apj, 553, L153
\bibitem[Hawarden \& Brand(1976)]{haw76} Hawarden, T.G., \& Brand, P.W.J.L. 1976, 
        \mnras, 175, 19
\bibitem[Henning \& Launhardt(1998)]{hen98} Henning Th., \& Launhardt R., 1998, A\&A 338, 223
\bibitem[Herbig \& Bell(1988)]{her88} Herbig, G. H., \& Bell, K. R. 1988, Lick Obs. 
        Bull. 1111
\bibitem[Hester et al.(1996)]{hes96} Hester, J. J., \etal 1996, \aj, 111, 2349
Merrill, K. M.,Gatley, I., Makidon, R. B., Meyer, M. R., \& Skrutskie, M. F. 1998, \aj, 116, 1816
\bibitem[Hillenbrand(2003)]{hil03} Hillenbrand, L.A. 2003,  
{\it Origins 2002: The Heavy Element Trail from Galaxies to Habitable Worlds}, 
eds. Woodward, C.E., \&  Smith, E.P., (ASP Conf. Ser.), astro-ph/0210520
        \aap, 302, 861 
\bibitem[Hoogerwerf et al.(2001)]{hoo01} Hoogerwerf, R., de Bruijne, J. H. J., 
        \& de Zeeuw, P. T. 2001, \aap, 365, 49
\bibitem[Keene et al.(1983)]{kee83} Keene, J., Davidson, J. A., Harper, D. A., 
        Hildebrand, R. H., Jaffe, D. T., Loewenstein, R. F., Low, F. J., \& 
        Pernic, R. 1983, \apjl, 274, 43
\bibitem[Kenyon \& Harmann(1995)]{ken95} Kenyon, S. J., \&  Hartmann, L. 1995, 
        \apjs, 101, 117
\bibitem[Khanzadyan et al.(2002)]{kha02} Khanzadyan, T., Smith, M. D., Gredel, R., 
        Stanke, T., \& Davis, C. J. 2002, \aap, 383, 502
\bibitem[Kim(2002)]{kim02} Kim, J. S. 2002, Ph.D. thesis, State University of New York, Stony Brook, NY
\bibitem[Kim et al.(2003)]{kim03a} Kim, J. S., Walter, F. M., \& Wolk, S. J. 2003,
         In Galactic Star Formation Across the Stellar Mass Spectrum, ASP Conference Series, 
         Vol. 287, proceedings of the 2002 International Astronomical Observatories in 
         Chile workshop, held 11-15 March 2002 at La Serena, Chile. Edited by James M. 
         De Buizer and Nicole S. van der Bliek. San Francisco: Astronomical Society of 
         the Pacific, ISBN: 1-58381-130-3, 2003, p. 275
\bibitem[King(1993)]{king93} King, J. R. 1993, \aj, 105, 1087
\bibitem[Knude et al.(1999)]{knu99} Knude, J., J{\o}nch-S{\o}rensen, H., \& Nielsen, 
        A. S. 1999, \aap, 350, 985
\bibitem[Knude \& Nielsen(2000)]{knu00} Knude, J., \& Nielsen, A. S. 2000, \aap, 
        362, 1138
\bibitem[Landolt(1992)]{lan92} Landolt, A. U. 1992, \aj, 104, 340
\bibitem[Lang(1980)]{lan80} Lang, K.R. 1980, Astrophysical Formulae,
         Springer-Verlag Berlin Heidelberg New York
\bibitem[Lefloch \& Lazaref(1994)]{lef94} Lefloch, B., \& Lazareff, B. 1994, 
        \aap, 289, 559
\bibitem[Lefloch \& Lazaref(1995)]{lef95} Lefloch, B., \& Lazareff, B. 1995, 
        \aap, 301, 522
\bibitem[Lefloch et al.(1997)]{lef97} Lefloch, B., Lazareff, B., \& Castets, A. 
        1997, \aap, 324, 249
\bibitem[Leggett et al.(2001)]{leg01} Leggett. S. K., et al. 2001, \apj, 548, 908
\bibitem[Luhman(1999)]{luh99} Luhman, K. 1999,  \apj, 525, 466
        2001, LPI, {\bf 32}, 1041
\bibitem[Mart{\'i}n et al.(1994)]{mar94} Mart{\'i}n, E.L., Rebolo, R., Magazzu, A., \& 
        Pavlenko, Ya. V. 1994, \aap, 282, 503
\bibitem[Mayor \& Mermilliod(1984)]{may84} Mayor, M. \& Mermilliod, J. C. 1984, 
        IAUS, 105, 411
\bibitem[Megeath et al.(1996)]{meg96} Megeath, S. T., Cox, P., Bronfman, L., 
        \& Roelfsema, P. R. 1996, \aap, 305, 296
\bibitem[Meyer et al.(1997)]{mey97} Meyer, M.R., Calvet, N., \& Hillenbrand, L.A. 
        1997, \aj, 114, 288 
\bibitem[Mink(1999)]{min99} Mink, D. J. 1999 in Astronomical Data Analysis Software and Systems VIII, ASP Conference Series, Vol. 172. Ed. David M. Mehringer, Raymond L. Plante, and Douglas A. Roberts. ISBN: 1-886733-94-5 (1999), p. 498
\bibitem[Monet et al.(1998)]{mon98} Monet, D., Bird A., Canzian, B., et al. 1998, 
         The USNO-A2.0 Catalogue, (U.S. Naval Observatory, Washington DC) 
\bibitem[Montmerle et al.(1983)]{mon83} Montmerle, T., Koch-Miramond, L.,
        Falgarone, E., \& Grindlay, J. 1983, ApJ, 269, 182.
\bibitem[Nielsen et al.(1998)]{nie98} Nielsen, A. S., Olberg, M., \& Booth, R. S.  
        1998, \aap, 336, 329
\bibitem[Nielsen et al.(2000)]{nie00} Nielsen, A. S., J{\o}nch-S{\o}rensen, H., \& Knude, J.
        2000, \aap, 358, 1077
\bibitem[O'Dell et al.(1993)]{odel93} O'Dell, C.R., Wen, Z., \& Hu., X. 1993, \apj, 
        410, 996
\bibitem[Oort \& Spitzer(1955)]{oort55} Oort, J.H., \& Spitzer, Jr. L. 1955, \apj, 
        121, 6
\bibitem[Pavlenko \& Magazzu(1996)]{pav96} Pavlenko, Y.V., \&  Magazzu, A. 1996, 
        \aap, 311, 961
\bibitem[Pettersson(1984)]{pet84} Pettersson, B. 1984, \aap, 139, 135
\bibitem[Pettersson(1987)]{pet87} Pettersson, B. 1987, \aap, 171, 101
\bibitem[Pettersson \& Reipurth(1994)]{pet94} Pettersson, B., \& Reipurth B. 1994, 
        \aaps, 104, 233
\bibitem[Pozzo et al.(2000)]{poz00} Pozzo, M., Jeffries, R.D., Naylor, T., 
        Totten, E.J., Harmer, S., \& Kenyon M. 2000, MNRAS {\bf 313}, 2310
\bibitem[Rao and Lambert(1993)]{rao93}Rao, K. N., \& Lambert, D. L. 1993, AJ, 105, 1915
\bibitem[Reipurth(1983)]{rei83} Reipurth, B. 1983, \aap, 117, 183 
\bibitem[Reipurth \& Zinnecker(1993)]{rei93} Reipurth, B., \& Zinnecker, H. 1993, 
        \aap,  278, 81
\bibitem[Reipurth et al.(2000)]{rei00} Reipurth, B., Yu, K. C., Heathcote, S.,
        Bally, J., \& Rodr{\'í}guez, L. F. 2000, \aj, 120, 1449
\bibitem[Reynolds(1976a)]{rey76a} Reynolds, R.J. 1976a, \apj, 203, 151
\bibitem[Reynolds(1976b)]{rey76b} Reynolds, R.J. 1976b, \apj, 206, 679
        Lewis, R. S., \& Grossman, L. 1998, Nature, 391, 559S
\bibitem[Sahu(1992)]{sah92} Sahu, M., 1992, Ph.D. Thesis, Groningen Univ.
\bibitem[Sahu \& Sahu(1993)]{sah93} Sahu, M., \& Sahu, K.C. 1993, \aap, 280, 231
\bibitem[Sandqvist(1976)]{san76} Sandqvist, A. 1976, \mnras, 177, 69
\bibitem[Schwartz(1977)]{sch77} Schwartz, R.D. 1977, \apj, 212, L25
\bibitem[Schwartz et al.(1990)]{sch90} Schwartz, R. D.,  Persson, S. E., \& 
        Hamann, F. W. 1990, \aj, 100, 793
\bibitem[Sherry et al.(2004)]{sher04}Sherry, W.H., Walter, F.M., \& Wolk, S.J. 2004, to be published in AJ, astro-ph/0410244
\bibitem[Sridharan(1992)]{sri92} Sridharan, T.K. 1992, J\aap, 13, 217 
\bibitem[Strom et al.(1989)]{str89} Strom, K.M., Wilkin, F.P., Strom, S.E., 
        \& Seaman, R. L. 1989, \aj, 98, 1444 
\bibitem[Sugitani et al.(1986)]{sug86} Sugitani, K., Fukui, Y., Ogawa, H., 
        \& Kawabata, K. 1986, \apj, 303, 667 
\bibitem[Sugitani et al.(1995)]{sug95} Sugitani, K., Tamura, M., \& Ogura, K. 1995, 
        \apjl, 455, 39
\bibitem[SAO(1997)]{ros97} U.S. ROSAT science data center/SAO 1997, 
        http://hea-www.harvard.edu/\\rosat/rsdc\_www/HRI\_CAL\_REPORT/hri.html
\bibitem[Vanhala at al.(1996)]{van96} Vanhala, H., Cameron, A.G.W., \&  Hoflich, P. 
        1996, LPI, 27, 1357
\bibitem[Ventura \& Zappieri(1998)]{ven98} Ventura, P. \& Zappieri, A.  1998, \aap, 
        340, 77
\bibitem[Walter et al.(1988)]{wal88} Walter F.M., Brown, A., Mathieu, R.D., 
        Myers, P. C., \& Vrba, F. J. 1988, \aj, 96, 297 
\bibitem[Walter et al.(1994)]{wal94} Walter F.M., Vrba, F.J., Mathieu, R.D., Brown, A., 
\& Myers, P.C. 1994, \aj, 107, 692
        \& Schmitt, J. H. M. M. 1997a, MmSAI, 68, 1081
\bibitem[Walter et al.(1997)]{wal97b} Walter, F.M., Vrba, F.J., Wolk, S.J.,
        Mathieu, R.D., Neuhauser, R. 1997, \aj, 114, 1544
\bibitem[Walter(2000)]{wal00} Walter, F.M. 2000, 
http://sbast3.ess.sunysb.edu/fwalter/CIRIM/cirim.html
\bibitem[Wolk \& Walter(1998)]{wol98} Wolk, S.J., \& Walter, F.M. 1997, in 
        {\it Cool Stars, Stellar Systems, and the Sun 10th}, eds. R. Donahue \& 
        J. Bookbinder (ASP: San Francisco), 1800  
\bibitem[Zealey et al.(1983)]{zea83} Zealey, W. J., Ninkov, Z., Rice, E., Hartley, 
        M., \& Tritton, S. B. 1983, \apjl, 23, 119
%
\end{thebibliography}
\end{document}